\documentclass[12pt]{article}

\usepackage{booktabs}
\usepackage{array,xcolor,colortbl}
\usepackage{multirow, makecell, colortbl, hhline}
\usepackage{cellspace}
\usepackage{verbatim}
\usepackage{cmap}
\usepackage{amsfonts}
\usepackage{amsmath}
\usepackage{amssymb}
\usepackage{amstext}
\usepackage{appendix}
\usepackage{hyphenat}
\usepackage{enumerate}
\usepackage{geometry}
\usepackage{graphicx}
\usepackage{setspace}
\usepackage{theorem}
\usepackage{color}
\usepackage{graphics}
\usepackage{graphicx}
\usepackage{latexsym}
\usepackage{natbib}

\usepackage{caption}
\usepackage{subcaption}
\usepackage{tikz}
\usepackage{times}
\usepackage{float}
\usepackage{fancyhdr}
\usepackage{tikz}
\usetikzlibrary{decorations.pathreplacing,calc,patterns}

\usetikzlibrary{arrows}
\usetikzlibrary{decorations.pathreplacing}

\newtheorem{corollary}{Corollary}
\newtheorem{definition}{Definition}
\newtheorem{lemma}{Lemma}
\newtheorem{proposition}{Proposition}

\renewcommand\refname{Bibliography}
\newcommand{\ba}{\begin{align}}
\newcommand{\ea}{\end{align}}
\newcommand{\be}{\begin{equation}}
\newcommand{\ee}{\end{equation}}
\newcommand{\bea}{\begin{eqnarray}}
\newcommand{\eea}{\end{eqnarray}}
\newcommand{\bean}{\begin{eqnarray*}}
\newcommand{\eean}{\end{eqnarray*}}
\newcommand{\ben}{\begin{enumerate}}
\newcommand{\een}{\end{enumerate}}

\newcommand{\bi}{\begin{itemize}}
\newcommand{\ei}{\end{itemize}}

\newcommand{\Var}{\mathrm{Var}}

\title{{\sc Truth and Deception}}

\date{}

\author{}

\pdftrailerid{}
\begin{document}

\author{\\{\sc Jeremy Bertomeu}\\ Washington University in St. Louis \\ \and \\ {\sc Edwige Cheynel}\\ Washington University in St. Louis\footnote{Jeremy Bertomeu is a professor and Edwige Cheynel is an associate professor at Washington University in St Louis. Contact author: Edwige Cheynel echeynel@wustl.edu. The authors thank John Barrios, Anne Beyer, Phil Dybvig, Richard Frankel, Keri Hu, Miles Geizmann, Christian Laux (discussant), Pierre Liang, Ivan Marinovic,  Xiumin Martin, Naomi Rothenberg, Bernard Sinclair-Desgagne (discussant), Jack Stecher, Kevin Smith, seminar participants at Stanford for helpful discussions, participants at EIASM Conference in Vienna, at the Bocconi University, the AES meeting in New York, and the University of Alberta. The methods are not based on experimental data and are exempt from pre-registration. Edited by Eduardo Azevedo.}}

\maketitle
\thispagestyle{fancy}
\begin{abstract}
Organizations rely on languages that imperfectly represent underlying states. We study the interaction of a lying cost, increasing in the distance from the message prescribed by the language, and a deception cost, increasing in the distortion induced in beliefs. Higher lying costs reduce bias and informativeness; higher deception costs do the opposite. Full revelation is sustained as lying costs become small, but with unbounded bias and communication costs. The most efficient policies sustaining nearly full revelation feature high deception costs with some tolerance for lying. With complementarities, targeting messages both untruthful and deceptive sustains full revelation without communication costs.
\end{abstract}

\vspace{0.5cm}



\newpage
\begin{singlespace}

\section{Introduction}

\paragraph{}

Should organizations penalize literal untruths or, instead, focus on whether a statement actually misled its recipient? This question has been widely debated in moral philosophy and law (\citealt{kan97,car10}), but its economic implications remain largely unresolved. On the one hand, language is inherently imperfect: words map reality into a limited set of messages and these messages cannot capture the full richness of the underlying state. Strict penalties for literal untruths may thus punish statements that convey relevant features more effectively than a literal description. For example, an exaggeration may convey a qualitative or subjective attribute that is otherwise difficult to express. On the other hand, an alternative perspective evaluates communication solely by its consequences on posterior beliefs and actions. From this viewpoint, what matters is whether the receiver would have taken a different action if fully informed; a literally true message may mislead a receiver expecting a biased message, while an untrue statement may be correctly understood once the bias is removed. For these reasons, literal truthfulness and accurate interpretations need not coincide. 

This study examines strategic communication when the sender bears a separable disutility from lying and deception. The sender has private information about the underlying state but must communicate using a language whose mapping into states is imperfect, because the available words may not capture all features of the underlying state relevant to decision-makers. We adopt the terminology of ``lying" as used in economics: to make a statement that is \textit{untruthful} relative to a specified language whose messages are anchored to the underlying state.\footnote{This follows the definition of lying in the lying-cost literature (\citealt{hurkar09,sob20}), although common usage is more equivocal. The Merriam-Webster dictionary offers two definitions where untruthfulness may or may not be sufficient for a lie, namely, (i) ``to make an untrue statement with intent to deceive'' or (ii) ``to create a false or misleading impression.'' In our setting, the underlying state also enters directly in the deception cost, so one could use an alternate definition for deception as lying relative to an accurate interpretation of the message. To avoid confusion, however, we reserve the term lying solely for untruthfulness.} Formally, a lie is captured by the distance between the chosen message and the message prescribed by the language.
  In contrast,  a message that induces beliefs  about the state that differ from the sender's expectation is \textit{deceptive}.
A message may  be literally untruthful but non-deceptive, because a predictable bias cannot deceive a rational receiver   \citep{ste89},
or literally truthful but deceptive, if the receiver expects bias and calibrates their interpretation accordingly.\footnote{Experimental evidence shows that individuals are willing to forgo material payoffs to avoid untruthful or deceptive communication \citep[e.g.,][]{gne05,karottsqu07,hurkar09,gnerocser13,gnekajsob18}.} We use the model to formally examine how the weights assigned to lying and deception affect bias, the quality of communication, and the signalling costs borne by the sender. Then, we discuss how systems should ideally design penalties to improve communication while keeping these costs low. 

The model captures the concepts of \textit{locutionary} falsity, referring to truth relative to commonly accepted meanings, and \textit{illocutionary} deception, in the sense of inducing false beliefs regardless of the message sent (\citealt{sob20}). In practice, while this distinction is neither easy to draw nor always successful in application, legal systems commonly attempt to distinguish objective falsity from misleading inference. In U.S. securities law, the SEC can enforce penalties when a public company materially misstates, even without evidence of investor reliance in decision-making. Damages awarded at trial are, however, typically tied to loss causation. Perjury and false-statements statutes can likewise turn on literal falsehood, without requiring proof of successful deception. Consumer-protection laws also recognize that a statement can be misleading through omission or selective emphasis, even without a false statement. This dual approach motivates us to treat untruthfulness and deception as separate components with potentially different implications. In particular, we view lying costs as capturing formal enforcement and other explicit penalties following an inaccurate report, such as fines for misstated financial information or perjury sanctions. This type of enforcement can be applied by comparing messages to an objectively shared language. The deception cost, by contrast, requires a more complex assessment of decision-making and can capture psychological or relational concerns, the process of determining loss causation, and, more generally, institutional features that align the sender's preferences with the receiver's, such as common goals or joint ownership.

Our main intuitions and results are summarized next. First, the sender's equilibrium report is either untruthful within the language or deceptive by inducing incorrect posterior beliefs. Second, lying and deception costs have opposing effects on communication. A higher lying cost reduces the sender's  bias but elicits the transmission of more language noise, reducing the precision of the receiver's belief. A higher deception cost has the opposite effect: it induces the sender to filter out language noise and makes the message more informative, but is accompanied with more predictable bias. Third, we study the sender's welfare and show that, for moderate deception costs, the sender prefers higher lying costs because they reduce costly bias; by contrast, increases in the deception cost unambiguously hurt the sender. 

We then ask whether certain penalty structures may be preferable, considering both their effect on communication and the costs borne by the sender. Paradoxically, communication becomes fully revealing when lying costs become infinitesimally small (while remaining nonzero), for any deception cost and any amount of language noise. Put differently, even a small lying cost tied to a language of potentially poor quality can restore full communication, although full revelation is impossible when lying costs are exactly zero. This outcome, however, binds the sender to unbounded disutility, because the equilibrium involves unbounded bias.
In practice, a sender placed in this situation would likely refuse to communicate if given an exit option. We therefore examine alternative designs that maintain large deception costs and finite lying costs. The optimal lying cost is higher when the sender's private information is more precise or when ex ante uncertainty is greater. Intuitively, when there is more meaningful information to communicate, the sender benefits from restraining incentives to misreport with lying costs.

In many realistic environments, lying and deception are interconnected rather than fully independent. Nevertheless, we begin with a specification that treats them as separate, so as to conceptually examine how strengthening one may affect communication and welfare. Put differently, the separable specification removes a built-in interaction that would otherwise be imposed in preferences rather than derived from strategic interaction. We then extend the model by introducing complementarities that increase marginal penalties when a message is both untruthful and deceptive. In a benchmark with only complementarities, the sender transmits exactly half of the language noise and, therefore, communication is higher than in a model with only lying costs, where the sender transmits all of the language noise. In this case, strengthening the complementarity does not affect communication but does reduce bias. When complementarities are combined with separable lying and deception costs, enforcement of the complementarity pushes the model toward this pure-complementarity outcome and generally has ambiguous effects on bias and communication, depending on whether lying costs or deception costs initially dominate. However, we also show that complementarities can be designed to reduce the costs borne by the sender. With sufficiently large deception costs and a positive weight on the complementarity, the sender's ex-ante cost can be made small. This outcome can be achieved for any lying cost, although the required weight on complementarities increases with the lying cost, because complementarities alleviate lying penalties when messages are not deceptive.\footnote{In a supplementary online appendix, we examine an extension of the model in which the benefit to the sender is stochastic and unknown to the receiver, as in \citet{fisver00}. In contrast to the baseline model, lying costs now play an additional role by reducing the noise arising from the benefit and can increase communication, creating environments in which both sender and receiver may prefer positive lying costs.}

Our approach builds on \citet{sob20}, who formalizes  the distinction between  lying and deception in a unified framework and studies their properties in strategic settings, showing that lying can sometimes be welfare enhancing.  Rather than focusing on classification, we explicitly model a parametric tradeoff between the two margins and analyze how their relative weight shapes equilibrium communication and welfare. We depart in two specific ways. First, truthfulness in our model is defined relative to an imperfect language: the sender's private information cannot be fully represented in a verifiable message. There may not be enough words to fully describe an economic situation, reporting standards simplify complex transactions, or the receiver may be unable to process arbitrarily rich information, so lying is evaluated against a noisy linguistic benchmark rather than the full state. Second, we introduce explicit penalties for both lying and deception and study how varying their relative strength affects bias and residual uncertainty.

Our model relates to an extensive literature on cheap talk and costly misreporting. Classic cheap-talk models such as \citet{crasob82} and \citet{far93} abstract away from lying costs and instead derive the credibility of communication from a partial alignment of preferences between sender and receiver. This approach, however, does not allow for an analysis of the interaction between lying costs and preference alignment, which is the primary focus of our study.
At the other extreme of the literature, because lying costs are pervasive in regulatory environments, several studies with applied contexts focus specifically on lying costs while leaving aside deception costs \citep[e.g.,][]{ste89,fisver00,dyesri04,gutkadkan06,frakar19}. In these models, lying costs discipline misreporting but may fail to restore informative communication when the language is imperfect.

Within this literature, our paper is most closely related to the foundational work of \citet{karottsqu07} and \citet{kar09}, who first combine cheap talk with lying costs within the partitional framework of \citet{crasob82}. Our focus, however, differs along several key dimensions. First, we study the implications of enforcing lying versus deception costs, whereas these earlier studies primarily characterize optimal communication. To address this question, we introduce a noisy language, as in \cite{dyesri04} and special cases of the generalized framework in \cite{frakar19}, which provides a friction preventing trivial solutions with arbitrarily large lying costs.
Second, we allow messages to be unbounded, whereas prior studies restrict messages through a finite type space. This creates an implicit reference point in which full revelation could arise if the language were perfect (\citealt{ste89}). By contrast, when messages are bounded, the bound may prevent full revelation because the highest types do not have separating messages.
Third, we focus on linear rational expectations equilibria rather than partitional solutions. Such equilibria are intuitive and commonly used in the noisy misreporting literature, and we further show that they are stable under iterated best responses.\footnote{It remains theoretically unresolved whether more efficient equilibria combining partitional and continuous message structures are possible. However, we show numerically that there are always linear equilibria that achieve higher sender utility and greater communication than the cheap-talk partitional equilibrium without lying costs.}

\section{The Model}

\subsection{Model setup}

\paragraph{}

A privately informed sender communicates with a receiver using an imperfect language, characterized by a mapping that does not preserve all the features of the underlying state. While we keep the terminology intentionally abstract, the model intends to capture many possible applications, such as a manager conveying information to investors, a divisional leader reporting to headquarters, an expert providing advice to a decision-maker, or a firm filing with a regulator, among other examples.
The actual state of the world is $\theta\sim N(0,\sigma_\theta^2)$, and the sender privately observes two signals:
\begin{itemize}
\item[(i)] a noisy signal about the state, $s=\theta+\epsilon$, where $\epsilon\sim N(0,\sigma_\epsilon^2)$; and 
\item[(ii)] a language noise $x\sim N(m_x,\sigma_x^2)$ that captures a possible discrepancy between the available language and the private information.
\end{itemize}

The report $r=R(s,x)$ is chosen by the sender to maximize  a utility function, composed of three terms:
\begin{equation}\label{U11}
u(r;s,x)
= \underbrace{P(r)}_{\text{(i) benefit from higher belief}}
- \underbrace{\frac{c}{2} (r-s-x)^2}_{\text{(ii) lying cost}}
- \underbrace{\frac{e}{2}\big(P(r)-\mathbb{E}(\theta| s)\big)^2}_{\text{(iii) deception cost}}.
\end{equation}

Hereafter, we set  $P(r)$ equal to the receiver's posterior expectation $P(r)=\mathbb{E}[\theta| r]$, so that the leading term (i) embeds an incentive to increase the receiver's posterior expectation.\ Equivalently, this term can be interpreted as arising from the receiver choosing an action $a$ to minimize  $\mathbb{E}[(a-\theta)^2| r]$, while the sender prefers a higher action. Because the sender benefits from higher beliefs regardless of whether they match the true state, this term generates a misalignment of preferences between sender and receiver.
When applied to the context of capital markets, $P(r)$ is often interpreted as a ``pricing function'' capturing how investors respond to a financial report.

The second component (ii) is a net cost 
$c>0$ per squared unit of bias and is similar in spirit to the assumption of lying costs in \cite{kar09}, where the message is anchored to a literal language. As in \cite{dyesri04} and \cite{frakar19}, we assume that the language is imperfect. The friction enters preferences through the noise term 
$x$, which creates a wedge between the truthful message within the language and the message most appropriate for the underlying state. Thus, the model captures that language cannot fully convey all features of the true state. To illustrate further, in a professional resume, 
$s$ may represent the true abilities of a prospective employee, while $s+x$ may correspond to observable credentials such as grades, degrees, or employment history. Alternatively, $x$ can be interpreted as communication ability: for example, a better salesperson, a more persuasive politician, or a more effective writer may generate a more favorable signal for the same underlying fundamentals. 

Finally, the third term (iii) represents the deception cost in Sobel's (2020) belief-driven notion of deception. A large difference $|P(r)-\mathbb{E}(\theta|s)|$ implies that the receiver chooses the incorrect action. In turn, this may imply negative consequences if
the sender intends to remain in good business terms with the receiver,
works in the same organization
with performance metrics in common,
or experiences moral costs from misleading. In a sale, the receiver may face ex-post liabilities or may be retaining a residual stake in the firm, which lose value if the receiver makes bad investment decisions by relying on the message.

Using the terminology in \citet{sob20}, component (ii) captures the locutionary act, that is, falsity relative to the language, while component (iii) captures the illocutionary act, namely deception through the beliefs induced by the message. A strictly positive variance $\sigma_x^2>0$ for the language noise is crucial for the trade-off we study. If $\sigma_x^2=0$, then for sufficiently large $c$ an enforcing authority could induce the sender to report $s$ arbitrarily closely, making fully revealing communication trivial and collapsing the distinction between lying and deception. When $\sigma_x^2>0$, by contrast, enforcement can only penalize deviations from an imperfect language benchmark. Increasing $c$ therefore reduces bias but induces the sender to transmit more exogenous language noise. Because our objective is to analyze optimal enforcement, we explicitly introduce this friction to capture the idea that arbitrarily large penalties may be problematic. In many realistic settings, the language on which enforcement is based does not perfectly align with the underlying state.

We restrict attention to linear strategies and beliefs and impose rational expectations. Specifically, we consider beliefs of the form
\begin{equation}\label{linP}
P(r)=\beta_0+\beta_1 r,
\end{equation}
where $\beta_1$ is the belief sensitivity to the report $\partial P(r)/\partial r$. Given a conjectured reporting strategy $R(s,x)$, we will require that $P(r)$ coincide with the Bayesian posterior expectation:
\begin{equation}\label{U12}
P(r) = \mathbb{E}(\theta| R(s,x)=r).
\end{equation}
\begin{definition}\label{defi1}
A linear rational expectations equilibrium, or LREE, is a pair $(R,P)$ such that
\begin{enumerate}
\item[(i)] the sender's reporting strategy is linear, $R(s,x)=\alpha_0+\alpha_1 s+\alpha_2 x$, and maximizes $u(r;s,x)$ for each $(s,x)$ given the belief function $P$ in \eqref{linP}; and
\item[(ii)] the belief function $P$ coincides with the Bayesian posterior \eqref{U12} induced by $R$.
\end{enumerate}
\end{definition}

We focus on linear rational expectations equilibria, which arise naturally with Gaussian uncertainty and quadratic payoffs (e.g., \citealt{rad79}; \citealt{kyl85}) and have been extensively used in models featuring noisy signalling. To provide additional confidence in the reasonableness of this equilibrium, in the Appendix, we show that the LREE is the unique fixed point of the best-response operator within the class of linear pricing functions. As a special case, iterated best responses converge to this solution starting from either a credible $P_0(r)=r$ or a fully skeptical $P_0(r)=0$ prior.

In what follows, we simplify notations by defining $\lambda\equiv \frac{\sigma_\theta^2}{\sigma_\theta^2+\sigma_\epsilon^2}$, so that $\mathbb{E}(\theta|s)=\lambda s$ is the sender's belief about the state, and $\Delta \equiv \frac{\sigma_\theta^2+\sigma_\epsilon^2+\sigma_x^2}{\sigma_\theta^2}
= \frac{Var(s+x)}{Var(\theta)}$
captures  the ratio of the variance of an unbiased report to fundamental variance. Note that $\lambda \Delta>1$ as long as $\sigma_x^2>0$.

\subsection{Linear equilibrium}
\label{lin1}
\paragraph{}

We focus on equilibria with $\beta_1>0$ in which the sender makes a report increasing in the state of the world.\footnote{We defer a full discussion of equilibria with negative sensitivity, $\beta_1<0$, to a supplementary online appendix, as they feature the counterintuitive property that posterior beliefs decrease in response to higher messages.} Exploiting the optimal reporting strategy in part (i) of Definition \ref{defi1}, we derive the reporting strategy  $r=R(s,x)$ by taking the first-order condition of (\ref{U11}), 
\begin{equation}\label{LX}
\beta_1
- c\big(R(s,x)-s-x\big)
- e\beta_1\big(P(R(s,x))-\mathbb{E}(\theta|s)\big)=0,
\end{equation}
noting that the second-order condition $-c -e (\beta_1)^2<0$ guarantees that the solution to (\ref{LX}) is a maximum.
Reorganizing the above equation and solving for $R(s,x)$, the  strategy is a linear function of the signal $s$ and $x$. Developing the sender's belief, $\mathbb{E}(\theta|s)=\lambda s,$ the reporting strategy is then given by
\begin{equation}\label{N11}
R(s,x)
=
\underbrace{\frac{\beta_1(1-\beta_0 e)}{c+\beta_1^2 e}}_{\equiv\alpha_0}
+
\underbrace{\frac{c+\lambda\beta_1 e}{c+(\beta_1)^2 e}}_{\equiv\alpha_1}s
+
\underbrace{\frac{c}{c+(\beta_1)^2 e}}_{\equiv\alpha_2}x,
\end{equation}
where the coefficients $(\alpha_i)_{i=0}^2$ describe the sender's optimal reporting strategy for given beliefs.

To impose rational expectations (\ref{U12}), we rewrite the message $R(s,x)=r$ in (\ref{N11}) as a signal of the state $\theta$ plus noise, that is,
\[
\frac{r-\alpha_0-\alpha_2 m_x}{\alpha_1}
=
\theta+\epsilon+\frac{\alpha_2}{\alpha_1}(x-m_x),
\]
implying that the expected state is given by
\begin{align}
P(r)
&=
\frac{\sigma_\theta^2}{\sigma_\theta^2+\sigma_\epsilon^2+\sigma_x^2\alpha_2^2/\alpha_1^2}
\left(\frac{r-\alpha_0-\alpha_2 m_x}{\alpha_1}\right)=
\underbrace{-\frac{\alpha_0+\alpha_2 m_x}{\frac{\alpha_1}{\lambda}+\frac{\alpha_2^2}{\alpha_1}\frac{\sigma_x^2}{\sigma_\theta^2}}}_{=\beta_0}
+
\underbrace{\frac{1}{\frac{\alpha_1}{\lambda}+\frac{\alpha_2^2}{\alpha_1}\frac{\sigma_x^2}{\sigma_\theta^2}}}_{=\beta_1}
\, r .
\label{N12}
\end{align}
Reinjecting $\alpha_1$ and $\alpha_2$ from (\ref{N11}) into (\ref{N12}), and solving the resulting equation for the sensitivity yields the following quadratic equation:

\begin{equation}\label{ol1}
\Gamma(\beta_1)
\equiv -e(\beta_1)^2+\beta_1\bigg(e \lambda- c\Delta\bigg)+ c =0.
\end{equation}

 Solving equation \eqref{ol1} in $\beta_1^*$ and simplifying all other coefficients from (\ref{N11})-(\ref{N12}) in terms of  $\beta_1^*$  characterizes the linear rational expectation equilibrium.

\begin{proposition}\label{equi}
For any $c>0$, there exists a unique LREE with positive belief sensitivity $\beta_1^*>0$ with:
\begin{itemize}
\item[(i)] The receiver's belief is $P^*(r)=\beta_0^*+\beta_1^* r$ with $\beta_1^*\in (1/\Delta,\lambda)$ given by the unique positive root of \eqref{ol1}, and
\begin{equation}\label{b0explicit}
\beta_0^*=-\beta_1^*\bigg(\frac{\beta_1^*}{c}+m_x\bigg).
\end{equation}
\item[(ii)] The reporting strategy is $R^*(s,x)=\alpha_0^*+\alpha_1^* s+\alpha_2^* x$ with

\begin{align}
\alpha_0^*
= \beta_1^*\left(\frac{1}{c} + \frac{\beta_1^* e m_x}{c+(\beta_1^*)^2 e}\right),\hspace{.2cm}
\alpha_1^*
=
\frac{c+\lambda \beta_1^* e}{c+(\beta_1^*)^2 e},
\hspace{.2cm}\alpha_2^*
= \frac{c}{c+(\beta_1^*)^2 e}. \label{aes}
\end{align}
\end{itemize}
Further, if $c=0$, any LREE must feature  $P^*(r)=0$. 
\end{proposition}

Several special cases provide helpful benchmarks for the general model.
First, absent deception costs ($c>e=0$), the sender does not directly care about $s$ separately from $x$, because they only enter the preference jointly as $s+x$. Therefore, the optimal reporting  strategy
involves considering only the state as represented by the language $s+x$ and must imply $\alpha_1^*=\alpha_2^*=1$ such that the entire language noise $x$ is passed into the message.
The belief sensitivity $\beta_1^*=\sigma_\theta^2/(\sigma_\theta^2+\sigma_\epsilon^2+\sigma_x^2)=1/\Delta$ thus does not depend on $c$, since no source of noise is filtered out by the sender, and a higher $c$ solely reduces the bias.\ In summary, lying costs discipline misreporting but, by themselves, do not benefit information transmission.

Second, consider the polar opposite case in which lying costs are exactly zero ($c=0$) while maintaining $e>0$, so that the sender is disciplined only by the deception penalty. In this case, the expressions in \eqref{aes} are no longer well-defined because the message is no longer anchored to a signal that depends on the fundamentals $s+x$. A direct argument nevertheless proves that there is a unique LREE, given by $P^*(r)\equiv 0$, that is, an uninformative babbling equilibrium.
This does not imply that communication is impossible in equilibria other than a LREE; indeed, any informative communication must rely on partitional messages (\citealt{crasob82}) in which fully revealing outcomes generally do not exist. Hence, there is a discontinuity at $c=0$: while the knife-edge case forces babbling (within the class of LREE), an arbitrarily small but positive lying cost $c>0$ restores a well-defined linear best response and can support substantial information transmission.\footnote{Indeed, we later prove that the LREE becomes fully informative when lying costs are small but strictly positive.} 

We expand next on the environment with zero lying costs by considering partitional equilibria, so that we may compare their informational and welfare properties to the baseline LREE. In this case, the underlying structure can be mapped into a special case of \citet{crasob82}, because the sender has separable preferences over the receiver's posterior belief and a quadratic loss from inducing beliefs that differ from her own expectation, while the receiver has a quadratic match-the-state preference.\footnote{Using notation closer to Crawford and Sobel, the sender's payoff can be written as
$U^{S}(P,r;\theta)=P-\frac{e}{2}\big(P-\mathbb{E}[\theta| r]\big)^{2}$,
while the receiver chooses an action $P$ to minimize
$\mathbb{E}\big[(P-\theta)^2| r\big]$, implying the optimal action
$P(r)=\mathbb{E}[\theta| r]$.
Equivalently, mapping Crawford and Sobel's ``action" $a$ to $P$, their state $m$ to $\theta$, and their message $y$ to $r$, the model corresponds to the classic framework with an alignment parameter proportional to $1/e$.}
As is standard in this class of models, fully separating regions do not exist. We therefore construct equilibria in which messages partition the signal space $s\in\mathbb{R}$ into intervals. Without loss of generality, the sender sends a message with label $r(s)=t_i$ whenever $s\in[t_i,t_{i+1})$, with cutoffs
$-\infty=t_0<t_1<\cdots<t_{N-1}<t_N=\infty$.
The message label $r(s)=t_i$ is unimportant and is written in terms of the message space only for expositional purposes. The belief is then given by
\[
 P(t_i)=\mathbb{E}[\theta| s\in[t_i,t_{i+1})]=\lambda\sigma_s\,\frac{\phi(t_i/\sqrt{\sigma_\theta^2+\sigma_\epsilon^2})-\phi(t_{i+1}/\sqrt{\sigma_\theta^2+\sigma_\epsilon^2})}{\Phi(t_{i+1}/\sqrt{\sigma_\theta^2+\sigma_\epsilon^2})-\Phi(t_i/\sqrt{\sigma_\theta^2+\sigma_\epsilon^2})},\]
where $\phi(.)$ and $\Phi(.)$ denote, respectively, the standard normal pdf and cdf. 

An $N$-bin partitional equilibrium is defined as in Definition \ref{defi1}, except that linearity is replaced by cutoff intervals. The indifference condition between two contiguous bins is
\[
\Big(P(t_{i-1})-\lambda t_i-\tfrac{1}{e}\Big)^2
=\Big(P(t_i)-\lambda t_i-\tfrac{1}{e}\Big)^2,
\]
which readily yields the recursion
\begin{equation}\label{cutoff_recursion}
\lambda t_i+\frac{1}{e}
=
\frac{P(t_{i-1})+P(t_i)}{2},
\qquad i=1,\dots,N-1,
\end{equation}
together with the boundary conditions $t_0=-\infty$ and $t_N=\infty$. For Gaussian signals, \citet{kazsargez21} establish that, for \textit{any} $N$, there exists a unique $N$-bin equilibrium and that equilibria with more bins are more informative.

When $c=0$, the sender faces no lying cost and therefore internalizes only the deception component of her objective. In the LREE, communication collapses to complete pooling (or one bin), so that the receiver remains uninformed and the sender bears the maximal expected deception loss. In comparison, any non-degenerate partitional equilibrium with more than one bin induces informative posteriors and strictly reduces this deception cost, thereby increasing both the sender's and the receiver's expected payoffs. Hence, at exactly $c=0$, any non-babbling partitional communication is preferable to the babbling LREE.
Importantly, this only holds absent lying costs and does not imply that a partitional equilibrium is preferred to every LREE with a non-zero cost vector $(c,e)$. We revisit this question later on after we fully further characterize bias and communication in the  LREE.
\section{Analysis}

\subsection{Preliminaries}

\paragraph{}

To better understand the implications of the linear equilibrium, we examine three constructs of interest that closely map to the penalties borne by the sender and the informativeness of the communication.
The first is the average bias of the report $r$ relative to the private signal $s$, given by
\begin{equation}\label{bias1}
B \equiv \mathbb{E}(r-s)
= -\frac{\beta_0^*}{\beta_1^*}
= \frac{\beta_1^*}{c}+m_x.
\end{equation}
Because the sender cannot increase prices in expectations, a higher bias is, all other things equal, a welfare loss to the sender, especially when lying costs per unit of bias are large. 

The second is the residual uncertainty or posterior variance of the state:
\begin{equation}\label{ie1}
\Sigma \equiv Var(\theta| r)
=
\sigma_\theta^2\left(
1-\frac{(\alpha_1^*)^2}{\frac{(\alpha_1^*)^2}{\lambda}+(\alpha_2^*)^2\left(\Delta-\frac{1}{\lambda}\right)}
\right)
=
\sigma_\theta^2(1-\beta_1^*\alpha_1^*).
\end{equation}

Under Gaussian distributions, where second moments serve as summary measures of uncertainty, the informativeness of the report can be measured as  the inverse of the residual uncertainty $\Sigma^{-1}$, which increases as the message conveys more information. Note that the residual uncertainty is also the expected utility of a receiver with a quadratic match-the-state preference, representing how more information translates into improved decision-making.

Finally, we define sender welfare $W$ as the expected utility of the sender. Because $\mathbb{E}(P(r))=\mathbb{E}(\theta)$ (by iterated expectations), the sender welfare is equal to the total expected  cost of the communication from the lying and deception components:
\begin{equation}\label{defi}
W
\equiv \mathbb{E}[u(r;s,x)] = 
-\underbrace{\mathbb{E}\bigg[\frac{c}{2}(r-s-x)^2\bigg]}_{w_c}
-\underbrace{\mathbb{E}\bigg[\frac{e}{2}\big(P(r)-\mathbb{E}(\theta| s)\big)^2\bigg]}_{w_e}.
\end{equation}

\subsection{Residual uncertainty and bias}

\paragraph{}

As the equilibrium of the model depends on the belief sensitivity, we first examine the comparative statics of the model in  $\beta_1^*$. We show below that deception costs increase this sensitivity, as can be verified from an application of the implicit function theorem on $\Gamma(.)$, \begin{equation}\label{lkx}
\frac{\partial \beta_1^*}{\partial e}
= \frac{\beta_1^*(\beta_1^*- \lambda)}{\Gamma'(\beta_1^*)}
>0,
\end{equation}
where the inequality follows from $\Gamma(.)$  decreasing at $\beta_1^*$ and $\beta_1^*\in (0,\lambda)$, as shown in Proposition \ref{equi}.
That the sensitivity should increase in response to greater deception costs is intuitive, but a complete explanation requires us to take a detour toward the receiver's residual uncertainty. 

The correct belief $\mathbb{E}(\theta|s)$ is a mean-preserving spread transformation of the (less informative)  belief $P(r)=\mathbb{E}(\theta|r)$, so that we can write the deception penalty using the  expression for $\beta_1^*$ in  \eqref{N12} as
 
\begin{align}
e\big(\mathbb{E}(\theta| r)-\mathbb{E}(\theta| s)\big)^2
&= e\,\mathrm{Var}\big(\mathbb{E}(\theta| s)| r\big) = e\,\big(\Sigma - \mathrm{Var}(\theta| s)\big).\label{hap1}
\end{align}

The rightmost term $Var(\theta|s)$ is exogenous, implying that the total penalty is, up to a constant, proportional to the residual uncertainty $\Sigma=Var(\theta|r)$. The higher the cost $e$, the more the sender focuses on reducing uncertainty by choosing a more informative message.
In turn, 
the belief responds more to the message, as noted in (\ref{lkx}).
In  the limit when the deception cost $e$ is large, the sensitivity $\beta_1^*$ converges toward $\lambda$, representing the Bayesian weight of a fully informative message,
because the sender prefers not to transmit any supplementary noise (the sensitivity $\alpha_2^*$ to the noise term $x$ converges to zero).

\begin{corollary}\label{e12}
An increase in the deception cost $e$ decreases uncertainty $\Sigma$ and increases bias $B$.
\end{corollary}

In Corollary \ref{e12}, it may seem surprising that deception costs, which make the sender more willing to convey information without error, increase average bias.
However, this predictable bias does not distort the receiver's belief and therefore, on its own, does not directly affect expected deception costs.  This logic is reflected in the bias term (\ref{bias1}), which is free from any term tied to deception costs except through the endogenous sensitivity $\beta_1^*$. However, deception costs indirectly affect bias via this belief-sensitivity channel. As the receiver expects more information to be conveyed in the message, sensitivity
increases, in turn magnifying incentives to bias.

Does the presence of deception cost fundamentally change the effect of the lying cost $c$ on communication? Recall that
in the benchmark of pure lying costs ($e=0$), the sender fully transmits the noise $x$ and therefore $c$ does not affect the belief sensitivity $\beta_1^*$. This is, however, no longer true with non-zero deception costs but the effect of lying costs is a-priori unclear: on the one hand, if deception costs incentivize the sender to remove some language noise ($\alpha_2^*<1$), the two costs may act as complements to offset the sender misalignment and implement reports closer to fundamentals; on the other hand, lying costs force the sender to transmit a greater fraction of the language noise. A comparative static exercise reveals that the second argument is correct, as
 lying costs always reduce belief sensitivity:\begin{equation}\label{betac}
\frac{\partial \beta_1^*}{\partial c}
= \frac{(\beta_1^*)^2 e\left(\beta_1^*-\lambda\right)}{c\left(c+(\beta_1^*)^2 e\right)}
<0.
\end{equation}
 
This observation has the following implications for the bias and residual uncertainty.

\begin{corollary}\label{e13}
An increase in lying costs $c$   increases residual uncertainty $\Sigma$ and decreases bias $B$. As $c \rightarrow 0$ (for any $e>0$ held constant),
the LREE becomes fully-revealing with $P(r)\rightarrow s$,  $\beta_1^* \rightarrow \lambda$ and  $\Sigma \rightarrow Var(\theta|s)$. 
\end{corollary}

Absent deception costs, the sender receives a benefit proportional to $c$
from filtering the measurement noise. This is the only added source of uncertainty and, therefore, the sender can only reduce uncertainty by controlling how measurement noise enters the receiver's belief, which also implies a benefit proportional to $e$. The trade-off is therefore obtained in terms of the ratio $e/c$. The greater this ratio, the more the sender de-noises the message, reducing $\alpha_2^*$.

As $c$ is reduced further toward zero, this logic implies the somewhat surprising observation that communication becomes fully revealing. This occurs for any level of language noise $\sigma_x^2$ and any fixed deception cost $e>0$, implying that even an arbitrarily small but positive lying cost restores perfect communication. Intuitively, the lying cost anchors the messages to a language that contains some information about the state, while keeping the cost small prevents the sender from transmitting the language noise. This is, of course, a limit result and should not be taken literally. Rather, our objective is to emphasize how even very small lying costs can dramatically alter equilibrium outcomes relative to the case in which lying costs are exactly zero. When $c=0$, perfect communication is no longer possible and any informative communication must rely on partitional strategies whose informativeness remains bounded away from full revelation. However, the bias becomes arbitrarily large as a lower lying cost incentivizes more bias and demonstrates that  weaker discipline on misreporting leads to increasingly extreme distortions in equilibrium messages.

\subsection{Sender welfare}

\paragraph{}

Improved communication benefits the receiver but may impose significant costs on the sender. For example, a large equilibrium bias may translate into substantial lying penalties, while deception costs may penalize reports that are truthful relative to the language but distort beliefs. One might therefore expect the sender to prefer low lying and deception costs. Indeed, in the extreme case of babbling with $e=c=0$, the sender achieves the highest possible surplus, $W=0$.
In this section we characterize the sender's preferences more generally and examine whether there exist combinations of $(c,e)$ that are preferred by both sender and receiver, or that can sustain perfect communication without imposing large costs on the sender.

Given that the sender's welfare $W=-w_c-w_e$ is the sum of the expected penalties from lying $-w_c$ and deception $-w_e$, it is convenient to re-express these terms in terms of characteristics of the equilibrium:  \begin{align}
w_c &= \frac{c}{2}\left(\mathrm{Var}(r-s-x)+(\mathbb{E}[r-s-x])^2\right)
= \frac{c}{2}\left(\mathrm{Var}(r-s-x)+\frac{(\beta_1^*)^2}{c^2}\right),\\
w_e &= \frac{e}{2}\big(\Sigma-\mathrm{Var}(\theta| s)\big).
\end{align}

The two additive terms in the expected lying cost $w_c$ capture separate channels through which the message may not conform with the language. The first term, $\mathrm{Var}(r-s-x)$, is the variance of the wedge between the report and the message that is truthful within the language, penalizing the sender for transmitting information that does not conform with the language (i.e., when $\alpha_2<1$). The second term $(\beta_1^*)^2/c^2$ measures the effect of the predictable bias and is increasing when the belief is more sensitive to the message.
Because $\beta_1^*$ is increasing in $e$, this term creates a link through which deception can increase expected penalties for lying.
The deception cost $w_e$ is simply a function of the reduction in residual uncertainty relative to the residual uncertainty of the sender.

\begin{proposition}\label{c2}
Let $w_c$ and $w_e$ denote the welfare loss due to lying and deception, as given in (\ref{defi}), 
\begin{itemize}
\item[(i)] as the lying cost $c$ increases, the lying loss $w_c$ decreases if $e$ is small and increases (resp., decreases) if $e$ is large and   $c^3\sigma_\theta^2(\lambda\Delta-1)>\lambda^3$ (resp., $c^3\sigma_\theta^2(\lambda\Delta-1)<\lambda^3$) holds. 
In contrast, the deception loss $w_e$ always increases in $c$.
\item[(ii)] as the deception cost $e$ increases, $w_c$ is increasing and $w_e$ is inverse U-shaped.
\end{itemize}
\end{proposition}

Without deception costs, a higher lying cost always reduces penalties because it forces the sender to reduce ex-post bias, which is ex-ante beneficial. This need not hold in the presence of deception costs. In general, $w_c$ is not monotone in $c$ and may increase with $c$ when the deception cost is sufficiently large.
To see why, note that there are two opposing forces. On the one hand, a higher $c$ induces the sender to lie less, which reduces the bias and therefore lowers the lying penalty. On the other hand, the deception cost $e$ pushes the sender to introduce predictable bias in order to influence beliefs, and each unit of bias becomes more costly when multiplied by $c$. When $e$ is sufficiently large, the report incorporates only little of the noise component $x$, so that most of the language noise is borne as bias relative to $s+x$. In this region, increasing $c$ raises the total cost of sustaining this bias, which explains why $w_c$ may increase with $c$.

The effect of $c$ on $w_e$, by contrast, is unambiguous and intuitive. A higher lying cost disciplines the sender to report closer to $s+x$ in order to avoid the lying penalty. This increases the amount of language noise transmitted to the receiver and therefore raises the deception loss $w_e$ through noisier beliefs.
We next consider the effect of changes in the deception cost $e$. Intuitively, a higher $e$ makes the sender more concerned with limiting belief distortions. This induces the sender to transmit more informative messages, which increases the bias component entering $w_c$ and therefore raises the lying loss.

At the same time, the deception loss itself behaves non-monotonically. When $e$ is close to zero, the total deception penalty $e(\mathrm{Var}(\theta|r)-\mathrm{Var}(\theta|s))$ remains small even though the expectation error is large, because the multiplicative factor $e$ is negligible. As $e$ increases from zero, the sender begins to bear higher deception penalties without substantially changing the informativeness of the message. In this region $w_e$ increases.

When $e$ becomes sufficiently large, however, the message becomes nearly fully informative so that $\mathrm{Var}(\theta|r)-\mathrm{Var}(\theta|s)$ becomes small. In the limit the sender’s reporting decision approximately satisfies
\[
P'(r) \approx e\,P'(r)\big(P(r)-\mathbb{E}(\theta|s)\big),
\]
which implies that the marginal belief distortion satisfies $P(r)-\mathbb{E}(\theta|s)\approx 1/e$. Hence the bias declines proportionally to $1/e$, and the total deception cost behaves as $e(1/e)^2=1/e$, which decreases in $e$. Intuitively, sufficiently large deception costs induce the sender to reduce deception, generating the inverse-U pattern for $w_e$.

To examine the total effect of lying and deception costs on the total sender's welfare, we establish below the following simplified characterization of welfare.

\begin{lemma}\label{lem} The equilibrium expected utility of the sender can be expressed as
\begin{eqnarray}\label{trl1}
W&=&-\frac{(\beta_1^*)^2}{2 c} -\frac{\sigma_\theta^2 e}{2   } \bigg( \lambda-\beta_1^*  \bigg).
\end{eqnarray}
\end{lemma}

Equation (\ref{trl1}) is useful not only for analyzing the comparative statics of sender welfare $W$, but also for providing intuition about the joint effects of lying and deception. To isolate the main forces, we first hold the belief sensitivity $\beta_1^*$ fixed. In the full comparative statics exercise, however, changes in costs also affect $\beta_1^*$ and must therefore be taken into account.
The first term, $-(\beta_1^*)^2/(2c)$, captures the effect of the lying cost. It is decreasing in $c$, reflecting the role of lying penalties: a higher lying cost induces the sender to reduce equilibrium bias. Although the cost per unit of bias increases, the resulting reduction in bias dominates, benefiting the sender. Deception costs affect this term only indirectly through their effect on $\beta_1^*$.
The second term, $-\sigma_\theta^2 e(\lambda-\beta_1^*)/2$, captures the deception penalty and symmetrically does not depend directly on the lying cost. This expression is proportional to the gap between the belief sensitivity under fully informative communication and the equilibrium belief sensitivity $\beta_1^*$. Intuitively, this term reflects the loss of information created by strategic reporting, since belief sensitivity and residual uncertainty are inversely related.

\begin{proposition}\label{P14}
Let the sender's expected utility $W$ be given by \eqref{trl1}:
\begin{itemize}
\item[(i)] $W$ is decreasing in $e$;
\item[(ii)] $W$ increases in $c$ if $e<\hat{e}\equiv\frac{1}{\sigma_x}\sqrt\frac{\Delta}{\lambda}$ and is inverse U-shaped in $c$ otherwise. As $c\to 0$ with $e>0$ fixed, the equilibrium bias $B$ diverges and the sender's expected penalties become unbounded.
\end{itemize}
\end{proposition}

Consider first the effect of deception costs.  Increasing $e$ \textit{always} reduces sender welfare. Although higher deception costs induce the sender to transmit more informative messages and reduce the deception component of the penalty, they simultaneously strengthen the belief response. This amplifies incentives to move beliefs through bias, raising the overall penalties borne by the sender. 

Changes in lying costs generate more complex interactions. When deception costs are small, increasing $c$ benefits the sender because reductions in bias dominate the associated increase in the per-unit cost of bias. When deception costs are larger, however, the sender filters part of the language noise in order to limit belief distortions, which increases the effective wedge between the report and the language. As a result, a higher lying cost increases the penalty associated with this wedge. When $c$ is initially small relative to $e$, the sender focuses primarily on reducing deception, so increasing $c$ raises the lying penalty and reduces welfare. As $c$ becomes larger, however, the sender transmits more of the language noise, bringing the equilibrium closer to the low-deception regime. Consequently, sender welfare becomes U-shaped in $c$ when deception costs are sufficiently large.

Proposition \ref{P14} suggests that the sender and receiver tend to have opposing preferences over the cost structure. The limiting case $c\to0$ is particularly telling. In this limit, the sender has no meaningful incentive to anchor communication to the language, creating the paradox that communication becomes fully informative even though the report seems no longer disciplined by the language. The mechanism sustaining this outcome is that the sender compensates for the weakening role of the language by introducing an arbitrarily large bias.
This results from two interacting effects. First, when lying costs are very small, the sender no longer transmits the language noise, so the message becomes fully informative. At the same time, the signalling value of the language becomes negligible, so an increasingly large bias is required to make the message credible. In the limit, this inefficient means of communication translates into unbounded losses for the sender.

This extreme result prompts our next question: can one design penalties that achieve fully revealing communication without imposing unbounded costs? The corollary below shows that the answer is positive and derives the optimal cost structure that, in the limit, implements a fully informative message preferred by the sender.

\begin{corollary}\label{JU}
Consider the set $\Omega$ of all sequences of penalty pairs $\omega=(c_i,e_i)_{i=1}^{\infty}$ such that, as $i\to\infty$, the associated residual uncertainty satisfies $\Sigma_i\to \mathrm{Var}(\theta| s)$ and the sender's welfare converges to $W_i \rightarrow W_\infty(\omega)$. Then, $\sup_{\omega \in \Omega} W_\infty(\omega) = -\lambda\sigma_x$ and the supremum can be approximated arbitrarily closely by sequences with $e_i\to \infty$ and $c_i\to c_\infty^{**}\equiv \lambda/\sigma_x$. \end{corollary}

To achieve a fully revealing message, the sender must either face small lying costs or large deception costs. The first alternatives is undesirable because it takes the form of unbounded bias, which leads to unbounded equilibrium penalties. The second alternative is therefore at least weakly preferable for the sender. Moreover, as shown in Proposition \ref{P14}, when deception costs are large the sender prefers an interior level of lying costs, implying that this alternative must allow some tolerance for lying. In fact, we show that there exists a combination of lying costs and sufficiently high deception costs that achieves fully revealing communication while keeping the sender's disutility bounded. Intuitively, the sender can control deception costs by filtering more language noise, but cannot eliminate a bias generated by equilibrium belief responses.

\begin{figure}[H]
\begin{center}
\includegraphics[width=\textwidth]{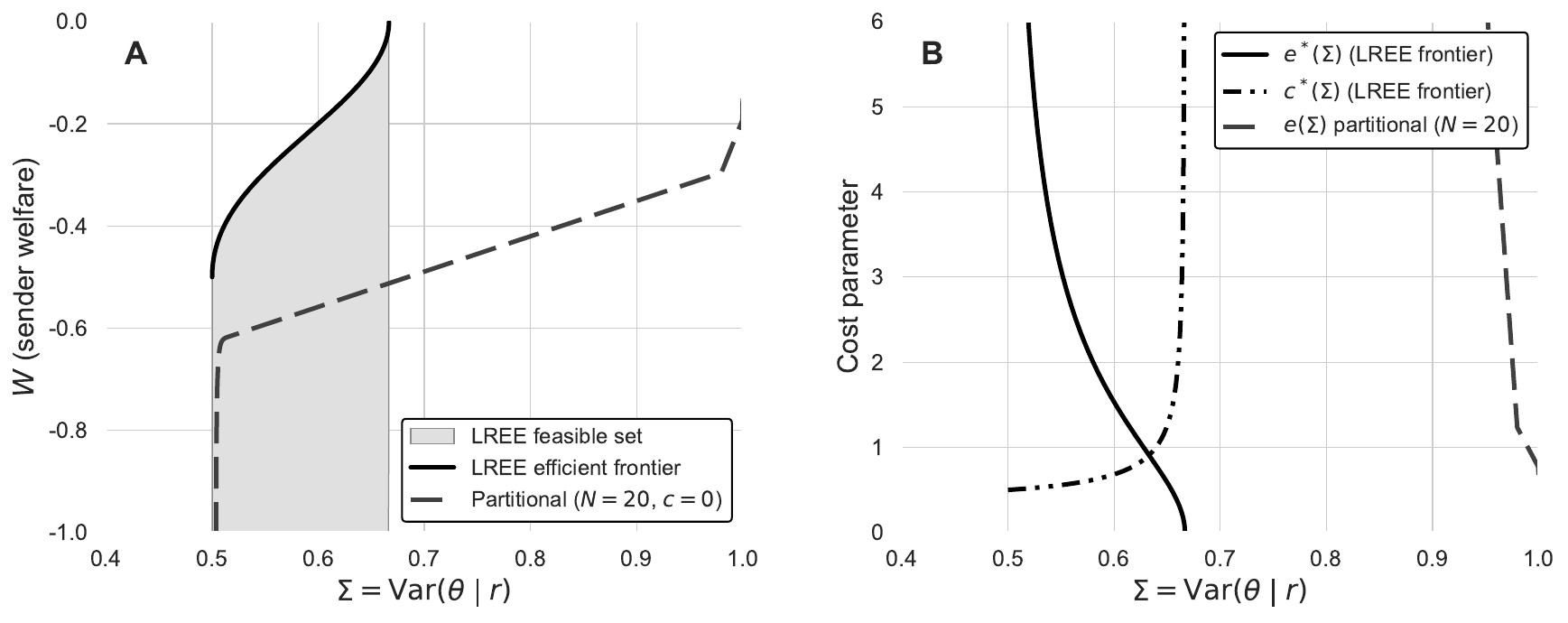}
\caption{Residual uncertainty, sender welfare and optimal costs $(e,c)$, with $\sigma_\theta=\sigma_\epsilon=\sigma_x=1$. Panel A: feasible pairs of residual uncertainty and sender welfare, with the efficient frontier and the partitional-equilibrium locus. Panel B: cost parameters implementing frontier outcomes. \label{by2}}

\vspace{0.3cm}
Alt text: Two line graphs. Panel A plots sender welfare against residual uncertainty, showing a shaded feasible region bounded above by an upward-sloping frontier and a dashed partitional-equilibrium curve lying below it. Panel B plots deception and lying cost parameters against residual uncertainty, showing one decreasing curve, one increasing curve, and a higher dashed partitional curve.
\end{center}
\end{figure}

In panel A of Figure \ref{by2}, we illustrate these observations in the $(\Sigma,W)$ plane of feasible residual uncertainty $\Sigma$ and sender welfare $W$. At one extreme, the point $(\Sigma,W)=(Var(\theta|s),W^{**})$ corresponds to the fully informative message characterized in Proposition \ref{JU}. This outcome is obtained when deception costs are large and the lying cost is set at $c_{\infty}^{**}$. Consistent with the trade-off between sender welfare and communication, the upper frontier of feasible $(\Sigma,W)$ pairs is upward sloping, indicating that the sender achieves higher welfare when communication is less informative. Movements along this frontier therefore represent alternative enforcement regimes that trade off communication quality against sender welfare.
Note that many policy combinations $(c,e)$ lie below this frontier, reflecting less effective enforcement. In this example, partitional equilibria reach below this frontier, implying that they are dominated by LREE; however, this property is not always true, as we shall discuss later on.
Panel B of the figure plots the set of optimal $(c,e)$ pairs that implement points along the frontier, demonstrating the conflict of preferences toward enforcement regimes: the sender prefers relatively higher lying costs, whereas the receiver prefers higher deception costs. As intuitive, enforcement using partitional equilibria and no lying costs require higher deception cost to achieve the same level of residual uncertainty.

We conclude this Section by comparing the LREE\ to the partitional equilibria with no lying cost $c=0$, in the limit case of fully-informative communication. As for the LREE, this requires in the partitional equilibrium that $e$ becomes large so that the partition becomes arbitrarily fine. As shown next, because the sender experiences in the limit unbounded penalty per fixed unit of deception, the sender's welfare does not converge to zero.

\begin{proposition}\label{PT_compare}
As $e\rightarrow \infty$, in the partitional equilibrium with a large number of bins, the message becomes fully-informative and the expected utility of the sender converges to $-\frac{\sigma_\theta}{2}\sqrt{\pi \lambda}$. In particular, sufficiently close to full-information, there exists a LREE\ with positive lying costs preferred by the sender to partitional equilibria if and only if $\sigma_\theta/\sigma_x>2\sqrt{\frac{\lambda}{\pi}}$, that is, when uncertainty is high or the language is less noisy.
\end{proposition}

Proposition \ref{PT_compare} offers some additional intuition as to when lying cost may be effective relative to no lying costs and partitional messages, in high information environment. The result is intuitive and suggests that frictions to language make a solution using lying costs less desirable to the sender; vice-versa, more ex-ante uncertainty lead to higher cost without using language as an anchor, increasing the benefit of lying costs.  

\section{Complementarity between Lying and Deception}
\paragraph{}
While cost separabilities allow us to examine how enforcement may target each type of cost, we recognize that many contexts treat lying and deception holistically, for example softening the marginal penalty on a lie if the lie could not conceivably deceive its recipient. Such complementarities are commonplace when multiple organizations focusing on each form of penalty interact
to evaluate the severity of a misrepresentation, say,  a lawsuit would be more likely to prevail on damages if it is preceded by an enforcement action noting a factual misrepresentation. This section accommodates the two costs jointly and shows that the main trade-offs apply in this richer setting. Specifically, we assume that\footnote{Because $r-s-x$ and $P(r)-\mathbb{E}(\theta| s)$ are unbounded under the Gaussian specification, the complementarity term may occasionally exceed the quadratic terms for some realizations, implying a negative realized sanction. We do not view this literally as a transfer scheme: a fixed rebate could improve welfare even without complementarities. Rather, $k$ is a reduced-form way to capture that non-deceptive lies should be penalized less severely than lies that also mislead. The results should therefore be interpreted conceptually: complementarities matter because they soften the penalties needed to sustain informative communication, not because they create literal subsidies. Later on, in analyzing welfare, we bound the parameters so that \textit{expected} penalties always dominate and, further, certain designs can make any transfer vanishingly small.}
\begin{equation}\label{U11_mult}
u(r;s,x)
= P(r)-\frac{c}{2}\,(r-s-x)^2-\frac{e}{2}
(P(r)-\mathbb{E}(\theta| s))^2-\frac{k}{2}\,(r-s-x)(P(r)-\mathbb{E}(\theta| s)).
\end{equation}
 Then, using steps similar to the baseline model, the equilibrium sensitivity $\beta_1^*$  is the unique positive root of the quadratic equation:

\begin{equation}\label{eq:Gamma_gen_restate}
\Gamma^c(\beta_1^*)
= \Gamma(\beta_1^*)
+k\frac{\lambda-(\beta_1^*)^2 \Delta }{2}
=0,
\end{equation}
which corresponds to the effect of separabilities $\Gamma(.)$ with a supplementary quadratic term capturing the effect of the complementarity. 

To set intuition, consider the benchmark of pure complementarities where $c,e\rightarrow 0$. Formally, this yields a benchmark belief sensitivity given only by the additional term in (\ref{eq:Gamma_gen_restate}):
\begin{equation}
\bar\beta_1^*=\frac{\sigma_\theta^2}{\sqrt{(\sigma_\theta^2+\sigma_\epsilon^2)\,(\sigma_\theta^2+\sigma_\epsilon^2+\sigma_x^2)}}=\sqrt{\frac{\lambda}{\Delta}}.
\end{equation}

Reinjecting the equilibrium condition into the sender's reporting rule, the coefficient on the language noise simplifies to $\alpha_2^*=\frac{1}{2}$, so that the sender filters exactly half of the language noise. That is, the sender symmetrically balances the message preferred for deception purposes, $r=s$, with the message that minimizes lying, $r=s+x$. The resulting level of communication does not depend on the magnitude of the complementarity, as in the case of pure lying costs. Yet, unlike this setting, the sender transmits a fraction of the language noise due to the presence of deception concerns. Consequently, any level of pure complementarities yields more informative communication than any level of pure lying costs. At the same time, because pure complementarities leave communication independent of $k$, they naturally raise the question of how complementarities interact with the separable penalties $(c,e)$, which we examine next.

\begin{proposition}\label{equi_genmix}
For any $c>0$, there exists a unique LREE with positive belief sensitivity $\beta_1^*>0$ with:
\begin{itemize}
\item[(i)] The receiver's belief is $P^*(r)=\beta_0^*+\beta_1^* r$ with $\beta_1^*\in (1/\Delta,\lambda)$ given by the unique positive root of (\ref{eq:Gamma_gen_restate}), and
\begin{equation}\label{b0explicit_gen}
\beta_0^*=-\beta_1^*\left(m_x+\frac{\beta_1^*}{c+\frac{k}{2}\beta_1^*}\right).
\end{equation}

\item[(ii)] The reporting strategy is $R^*(s,x)=\alpha_0^*+\alpha_1^* s+\alpha_2^* x$ with
\begin{align}
\alpha_2^*
=\frac{c+\frac{k}{2}\beta_1^*}{c+k\beta_1^*+e(\beta_1^*)^2},
\alpha_1^*
=\frac{c+\frac{k}{2}\beta_1^*+\lambda\left(e\beta_1^*+\frac{k}{2}\right)}{c+k\beta_1^*+e(\beta_1^*)^2},
\alpha_0^*
=\frac{\beta_1^*}{c+\frac{k}{2}\beta_1^*}
+\frac{\beta_1^*\left(e\beta_1^*+\frac{k}{2}\right)}{c+k\beta_1^*+e(\beta_1^*)^2}\,m_x.\raisetag{45pt}
\label{aes_gen}
\end{align}
\end{itemize}
Hence, $\beta_1^*$ is increasing in $e$ and decreasing in $c$, and varies in $k$ toward $\bar\beta_1^*$ (that is, is increasing in $k$ if $c>\frac{e\lambda}{\Delta}$ and decreasing in $k$ otherwise).
\end{proposition}

The next corollary characterizes how complementarities affect bias and residual uncertainty.

\begin{corollary}\label{sigma_bias_gen}
Residual uncertainty $\Sigma$ is decreasing in $e$ and increasing in $c$. Moreover, it varies with $k$ in the direction of the pure-complementarity benchmark: $\Sigma$ decreases when $\beta_1^*<\bar\beta_1^*$ and increases otherwise.
The average bias $B$ is decreasing in $c$, increasing in $e$, and decreasing in $k$ when $\beta_1^*\ge \bar\beta_1^*$ (but ambiguous in $k$ otherwise). \end{corollary}

The corollary confirms that complementarities act as a force shifting the equilibrium toward the pure-complementarity benchmark $\bar\beta_1^*$. If the separable model features too little sensitivity relative to $\bar\beta$, because lying costs dominate deception costs, stronger complementarities increase communication. The opposite occurs when belief sensitivity initially exceeds the benchmark. One may then derive a welfare expression analogous to Lemma~\ref{lem}, except that $k$ now enters both through the equilibrium sensitivity $\beta_1^*$ and through the
determinant $4ce-k^2$ of the sender's quadratic objective that captures the interaction between lying and deception costs.
\begin{corollary}\label{welfare_gen}
The sender welfare satisfies
\begin{equation}\label{eq:W_gen}
W \equiv \mathbb{E}[u(r;s,x)]
=
-\underbrace{\frac{c\,(\beta_1^*)^2}{2\big(c+\frac{k}{2}\beta_1^*\big)^2}}_{(A)}
-\underbrace{\frac{(4ce-k^2)\,\sigma_\theta^2}{8\big(c+\frac{k}{2}\beta_1^*\big)}(\lambda-\beta_1^*)}_{(B)}.
\end{equation}
\end{corollary}

Two terms emerge from \eqref{eq:W_gen}.
The first term (A) captures the expected lying cost. The sender distorts the report away from the message prescribed in the language $s+x$ in order to shift beliefs. The complementarity $k$, in the denominator, softens the lying penalty when the message is correctly processed. The second term (B) captures the welfare loss from deception. As before, this penalty vanishes when $\beta_1^*=\lambda$, that is, when the receiver responds as if they were directly observing the unmuddled signal $s$. The incremental role of complementarities operates through the factor $4ce-k^2$. As $k^2$ approaches the boundary $4ce$, the system becomes increasingly designed to shield the sender against deception that is not caused by a lie.

These considerations imply that while the sender's welfare can be preserved through the use of complementarities, although the complementarities may also damage the quality of communication by shifting the equilibrium toward a price response $\bar \beta_{1}^*<\lambda$. It is therefore unclear whether using $k$, jointly with $(c,e)$ would improve communication at lower cost to the sender. To resolve this problem, we restrict attention to regimes $k^2\le 4ce$, so that the joint penalty remains convex. This implies that the penalized component of utility is nonnegative, so that the sender's expected payoff cannot exceed $\mathbb{E}[\theta]=0$ and no ex-ante subsidy is needed.

\begin{corollary}\label{fullinfo_gen}
Consider the set $\Omega$ of all sequences of penalty triples $\omega=(c_i,e_i,k_{i})_{i=1}^{\infty}$ such that $k_i^2\le 4c_i e_i$ and, as $i\to\infty$, the associated residual uncertainty satisfies $\Sigma_i\to \mathrm{Var}(\theta| s)$ and the sender's welfare converges to $W_\infty(\omega)$. Then, $\sup_{\omega \in \Omega} W_\infty(\omega) = 0$ and this supremum can be approximated arbitrarily closely by sequences with $e_i\to \infty$, any fixed  $c_i\to c^0$, and $k_i \sim 2\sqrt{c^0 e_i}$.
 \end{corollary}

Complementarities can therefore allow the regulator to approach full revelation without communication losses to the sender.
By raising $e$ while keeping $k$ near the boundary $k^2\approx 4ce$, the regulator strengthens the incentive to align beliefs with fundamentals and simultaneously weakens the effective anchoring discipline of $c$.
In the limit $\beta_1^*\to\lambda$, so residual uncertainty falls to $\mathrm{Var}(\theta|s)$, and the misalignment component of \eqref{eq:W_gen} vanishes.
The lying distortion also vanishes because $c+\frac{k}{2}\beta_1^*$ diverges when $k\sim\sqrt{e}$, driving expected welfare to zero.
Lying does not become costless; rather, the marginal penalty for benchmark deviations shrinks relative to the growing incentive to align beliefs with the state.

\section{Conclusion}

\paragraph{}

Our study is the first to formally demonstrate that a joint consideration of lies and deception can improve communication without incremental costs to the sender.
We show that, in many settings, senders may prefer to face higher lying costs rather than deception penalties and that, if both penalties can be controlled, more effective communication should feature very high penalties for deception jointly with some tolerance for untruthfulness. We also demonstrate, perhaps surprisingly,  that when the lying cost is approaching zero, full revelation arises but the reporting bias is arbitrarily large. This result stands in sharp contrast with models that consider lying costs alone, where communication collapses as lying costs vanish, emphasizing that even minimal lying costs, when combined with deception concerns, can fundamentally restore and discipline  information transmission.

These results help explain why there may be significant disagreement about the moral penalties associated with lying, as well as why, within legal systems, the type of penalty often depends on the alleged offense. Societies with stronger institutions, which tend to exhibit greater expectations of trust, appear to rely less on strict literal interpretations of the law and more on assessing deception. Under common law, for example, deception is implicit in the assessment of whether a plaintiff acted in reliance on a statement. To give another example, an employer may rescind a promised job offer, or the offer may not have been serious in the first place, but a plaintiff typically cannot recover unless they took actions (such as abandoning their current job or moving) based on the belief that the offer would be honored. By contrast, in societies where low trust is expected, especially those with weaker institutions, punishments appear to rely more heavily on explicit penalties and strict applications of the law.

Our broader objective is to contribute to the question of how institutions can be designed to improve communication. Much of the existing literature on noisy signalling and cheap talk focuses on the form of communication, which we take here as given and therefore do not study directly. Instead, we examine the trade-off in using different types of penalties, offering new insights into how institutions can shape communication incentives. A key assumption in our setting is that language is imperfect, placing limits on what penalties for lying can achieve in principle. That said, important theoretical questions remain open regarding how noisy signalling and cheap talk may interact, since our focus on linear rational expectations equilibria excludes partitional communication structures. It remains unresolved whether equilibria outside the linear class may lead to more efficient communication.

Finally, although our focus is on communication, similar considerations arise in the broader debate between rules and principles. Communication within the rules of a language represents one class of rules, but many other rules prescribe behavior relative to a pre-specified action that may not be perfectly suited to every context. Principles, by contrast, offer greater flexibility, permitting deviations when additional considerations, such as welfare, are relevant and when the explicit codification of all contingencies is impractical. For example, although a posted speed limit imposes a maximum legal speed, adverse road conditions may require slower driving. Similarly, in emergencies, actions such as swimming in a restricted area may be justified to save a life. This dual approach to enforcement raises important questions in law and economics regarding the appropriate balance between rules and principles.

\section*{Appendix A: Main Text Omitted Proofs}

\textbf{Proof of Proposition \ref{equi}:}
For $c>0,$ we consider the (continuous) function $\Gamma(\beta_1)$ in \eqref{ol1}. Noting that $
\Gamma\left(\lambda\right)
=-c (1-\lambda \Delta)<0<\Gamma(1/\Delta)=
\frac{e}{\Delta^2}(\lambda\Delta-1)$,
the intermediate value theorem implies the existence of at least one root $\beta_1^*>0$ in $(1/\Delta,\lambda)$.  To prove uniqueness, it suffices to show that the derivative of the above function evaluated at $\beta_1^*$ is negative, which can be established by rewriting equation $\Gamma(\beta_1^*)=0$ as \begin{equation}
\label{sig}
-e\beta_1^*\left(2\beta_1^*-\lambda\right)-c\Delta\beta_1^*=-e(\beta_1^*)^{2}-c
\end{equation}
and substituting the above into the expression for $\Gamma'(\beta_1^*)$: 
\begin{equation}\label{der}
\Gamma'(\beta_1^*)
=
-e\left(2\beta_1^*-\lambda\right)
-c\left(
\frac{1}{\beta_1^*}
-\frac{\beta_1^* e}{c}
+\frac{e\lambda}{c}
\right)
=
-\frac{c+(\beta_1^*)^2 e}{\beta_1^*}
<0.
\end{equation}
The constant    $\beta_0^*$  is then obtained from (\ref{N11}) and (\ref{N12}) as 
\begin{eqnarray}\label{eq1}
\beta_0^*=-\beta_1 \bigg(\alpha_0 +\alpha_2 m_x\bigg)  =\frac{\beta^*_1 \bigg(\beta^*_1 ( \beta^*_0 e-1) - c m_x\bigg)}{(c + (\beta^*_1)^2 e)}.\end{eqnarray} Solving for equation \eqref{eq1},    $\beta_0^*$  is defined by $\beta_0^*=-\beta^*_1 (\frac{\beta^*_1}{c} + m_x)$ and the terms of the reporting strategy $(\alpha_0^*,\alpha_1^*,\alpha_2^*)$ follow readily from (\ref{N11}). 

Consider next the case  $e>c=0$ and suppose that there exists a linear REE with
$\beta_1>0$. Then \eqref{ol1} reduces to
$\Gamma(\beta_1)=e\beta_1\big(\lambda-\beta_1\big)$,
so necessarily $\beta_1=\lambda>0$.
However, the condition \eqref{eq1} with $c=0$ implies
$\beta_0^*=\beta_0^*-1/e$, a contradiction. Hence, no non-degenerate linear REE exists when $c=0$ and $e>0$.
An uninformative pooling equilibrium can be supported with constant beliefs $P(r)\equiv 0$
(i.e., $\beta_0^*=\beta_1^*=0$) and, for instance, the constant reporting strategy $R(s,x)\equiv 0$
(more generally, any reporting rule independent of $s$).$\Box$
\medskip\\\textbf{Proofs of Corollaries \ref{e12} and \ref{e13}:}
We derive first the comparative statics in $e$, starting with the residual uncertainty $\Sigma$. Substituting the expression for $\alpha_1^*$ in \eqref{aes} into \eqref{ie1} yields
\begin{equation*}
\Sigma=
\sigma_\theta^2\left(1-\beta_1^*\frac{c+\lambda \beta_1^* e}{c+(\beta_1^*)^2 e}\right).
\end{equation*}
Differentiating with respect to $e$,
\begin{equation}
\frac{\partial \Sigma }{\partial e}
=
\frac{c\sigma_\theta^2\left((\beta_1^*)^2(\beta_1^*-\lambda)
-\big(c+e(2\lambda-\beta_1^*)\beta_1^*\big)\frac{\partial \beta_1^*}{\partial e}\right)}
{\big(c+(\beta_1^*)^2 e\big)^2}
<0,
\label{var1}
\end{equation}
where the inequality follows from inserting expression of $\partial \beta_1^*/\partial e$ from \eqref{lkx} and the expression of $\Gamma'(\beta_1^*)$ from \eqref{der}. For the bias, differentiating with respect to $e$ readily yields $\frac{\partial B}{\partial e}
=\frac{1}{c}\frac{\partial \beta_1^*}{\partial e}>0$.

For the comparative statics in $c$, differentiating $\Sigma$ implies
\begin{align*}
\frac{\partial \Sigma }{\partial c}
&=
\frac{\sigma_\theta^2\left((\beta_1^*)^2 e\left(\lambda-\beta_1^*\right)
- c\big(c+e(2\lambda-\beta_1^*)\beta_1^*\big)\frac{\partial \beta_1^*}{\partial c}\right)}
{\big(c+(\beta_1^*)^2 e\big)^2}
>0
\end{align*}
and, from \eqref{bias1}, $
\frac{\partial B}{\partial c}
=\frac{1}{c^2}\left(c\frac{\partial \beta_1^*}{\partial c}-\beta_1^*\right)<0.$
The limit as $c\to 0$ follows from the expression of $\beta_1^*$. $\Box$
\medskip\\\textbf{Proof of Proposition \ref{c2}:} Since the proposition makes several claims, we break down its proof in separate parts.

\textit{Part 1: Welfare Expressions.} The expected lying penalty is
\[
w_c
\equiv
\mathbb{E}\!\left[\frac{c}{2}(r-s-x)^2\right]
=
\frac{c}{2}\left(Var(r-s-x)+\big(\mathbb{E}(r-s-x)\big)^2\right).
\]Simplifying the variance term:\\
\begin{align*}Var(r-s-x)
&=
(\alpha_1^*-1)^2(\sigma_\theta^2+\sigma_\epsilon^2)+(\alpha_2^*-1)^2\sigma_x^2.
=
\frac{(\beta_1^*)^2e^2\sigma_\theta^2}{\big(c+(\beta_1^*)^2e\big)^2}
\left[
\frac{(\beta_1^*-\lambda)^2}{\lambda}
+(\beta_1^*)^2\!\left(\Delta-\frac{1}{\lambda}\right)
\right]\\&=
\frac{e^2(\beta_1^*)^2\sigma_\theta^2(\lambda-\beta_1^*)}
{c\big(c+(\beta_1^*)^2e\big)},\end{align*}
where the second equality follows from substituting the equilibrium coefficients and
inserting $\sigma_x^2=\sigma_\theta^2\big(\Delta-\frac{1}{\lambda}\big)$ and the third equality uses \eqref{sig} to insert    $\Delta=\frac{1}{\beta_1^*}-\frac{\beta_1^*e}{c}+\frac{e\lambda}{c}$. 

The expectation simplifies to $\mathbb{E}(r-s-x)=m_x-(\alpha_0^*+\alpha_2^*m_x)=\frac{\beta_1^*}{c}$. Then, combining the variance and expectation terms returns
\[
w_c
=
\frac{c}{2}\left(Var(r-s-x)+\big(\mathbb{E}(r-s-x)\big)^2\right)=
\frac{c}{2}\left(
\frac{e^2(\beta_1^*)^2\sigma_\theta^2(\lambda-\beta_1^*)}
{c\big(c+(\beta_1^*)^2e\big)}
+\frac{(\beta_1^*)^2}{c^2}
\right).
\]
Similarly, 
the expected deception penalty simplifies to a variance:  \[w_e
\equiv
\mathbb{E}\!\left[\frac{e}{2}\big(P(r)-\mathbb{E}(\theta| s)\big)^2\right]
=
\frac{e}{2}\mathbb{E}\!\left[\big(\beta_0^*+\beta_1^*r-\lambda s\big)^2\right]=
\frac{e}{2}\Var\big(\beta_0^*+\beta_1^*r-\lambda s\big),\]
where the third equality follows from the mean satisfying
$\mathbb{E}\big(\beta_0^*+\beta_1^*r-\lambda s\big)=\beta_0^* + \beta_1^* (\alpha_0^* + \alpha_2^*  m_x)=0$ given that   $\beta_0^*=-\beta_1^*(\alpha_0^*+\alpha_2^*m_x)$. Substituting the equilibrium coefficients further yields
\[
Var\big(\beta_0^*+\beta_1^*r-\lambda s\big)
=
\frac{c^2}{\big(c+(\beta_1^*)^2e\big)^2}
\left[
\frac{(\beta_1^*-\lambda)^2}{\lambda}\sigma_\theta^2+(\beta_1^*)^2\sigma_x^2
\right]=\frac{c\sigma_\theta^2(\lambda-\beta_1^*)}{c+(\beta_1^*)^2e};
\]
after  substituting  $\sigma_x^2=\sigma_\theta^2(\Delta-\frac{1}{\lambda})$ and     $\Delta=\frac{1}{\beta_1^*}-\frac{\beta_1^*e}{c}+\frac{e\lambda}{c}$,  it follows that 
\begin{equation}\label{we_clean}
\hspace{.5cm}w_e
=
\frac{ec\sigma_\theta^2(\lambda-\beta_1^*)}
{2\big(c+(\beta_1^*)^2e\big)}.
\end{equation}
\textit{Part 2: Comparative statics in $c$.} Differentiating $w_c$,
 \begin{eqnarray}
\frac{\partial w_c}{\partial c}
&=
\frac{\partial \beta_1^*}{\partial c}
\left[
\frac{\beta_1^*}{c}
+
\frac{e^2\sigma_\theta^2\beta_1^*
\big(c(2\lambda-3\beta_1^*)-e(\beta_1^*)^3\big)}
{2\big(c+e(\beta_1^*)^2\big)^2}
\right]
-\frac{(\beta_1^*)^2}{2}
\left[
\frac{1}{c^2}
+
\frac{e^2\sigma_\theta^2(\lambda-\beta_1^*)}
{\big(c+e(\beta_1^*)^2\big)^2}
\right]\nonumber\\\label{eq}&=
-\frac{(\beta_1^*)^2}{2}
\left[
\frac{1}{c^2}
+\frac{2e\beta_1^*(\lambda-\beta_1^*)}{c\big(c+e(\beta_1^*)^2\big)}
+\frac{e^2\sigma_\theta^2(\lambda-\beta_1^*)}{\big(c+e(\beta_1^*)^2\big)^2}
\left(
1+\frac{e\beta_1^*\big(c(2\lambda-3\beta_1^*)-e(\beta_1^*)^3\big)}
{c+e(\beta_1^*)^2}
\right)
\right],
\end{eqnarray}
after substituting $\frac{\partial\beta_1^*}{\partial c}
=
\frac{(\beta_1^*)^2 e(\beta_1^*-\lambda)}
{c+(\beta_1^*)^2 e}$ given by \eqref{betac} and \eqref{der}.
Note that the sign of $\partial w_c/\partial c$ can be positive or negative but, since \(\beta_1^*\to1/\Delta\) as \(e\to 0\), $\partial w_c/\partial c\rightarrow -1/(2\Delta^2c^{2})<0$ is negative when $e$ is small enough. The sign of the comparative static when $e$ is large requires an additional condition, which we derive next. Since 
$\beta_1^*\to\lambda$ as $e\to\infty$, using \eqref{ol1},  it follows that
\[
\lim_{e\rightarrow \infty}e(\lambda-\beta_1^*)=
c\left(\Delta-\frac1\lambda\right)
=\frac{c(\lambda\Delta-1)}{\lambda}>0.
\]
Hence the second term in \eqref{eq} vanishes $\frac{2e\beta_1^*(\lambda-\beta_1^*)}{c\big(c+e(\beta_1^*)^2\big)}
\longrightarrow 0$. Next, the last term in \eqref{eq} has only vanishing terms except for the leading term which, from the limit above, converges to\[\lim_{e\rightarrow \infty}\frac{e^2\sigma_\theta^2(\lambda-\beta_1^*)}{\big(c+e(\beta_1^*)^2\big)^2}
\left(
1+\frac{e\beta_1^*\big(c(2\lambda-3\beta_1^*)-e(\beta_1^*)^3\big)}
{c+e(\beta_1^*)^2}
\right)=
\frac{c(\lambda\Delta-1)}{\lambda^3}.\] Regrouping these observations:
\[
\lim_{e\rightarrow \infty}\frac{\partial w_c}{\partial c}
=-\frac{\lambda^2}{2}
\left[
\frac1{c^2}
-\frac{c\sigma_\theta^2(\lambda\Delta-1)}{\lambda^3}
\right]
=
\frac12\left[
\frac{c\sigma_\theta^2(\lambda\Delta-1)}{\lambda}
-\frac{\lambda^2}{c^2}
\right],
\]
which has the sign of $c^3\sigma_\theta^2(\lambda\Delta-1)-\lambda^3$.

For $w_e$, differentiating \eqref{we_clean} with respect to $c$ and substituting $\partial\beta_1^*/\partial c$ from \eqref{betac} and \eqref{der} returns
\[
\frac{\partial w_e}{\partial c}
=
\frac{e^2\sigma_\theta^2(\beta_1^*)^2(\lambda-\beta_1^*)
\big(c+\lambda \beta_1^* e\big)}
{\big(c+(\beta_1^*)^2e\big)^3}
>0.
\]
\textit{Part 3: Comparative statics in $e$.} Using  \eqref{lkx} and \eqref{der} to insert $ 
\frac{\partial\beta_1^*}{\partial e}
=
\frac{(\beta_1^*)^2(\lambda-\beta_1^*)}
{c+(\beta_1^*)^2 e}$,
\\
\[
\frac{\partial w_{c}}{\partial e}
=
\frac{\partial \beta_1^*}{\partial e}
\left[
\frac{\beta_1^*}{c}
+
\frac{ce\,\sigma_\theta^2\big(c+\beta_1^* e\lambda\big)}
{\big(c+e(\beta_1^*)^2\big)^2}
\right]=
\frac{(\beta_1^*)^2(\lambda-\beta_1^*)}
{c+e(\beta_1^*)^2}
\left[
\frac{\beta_1^*}{c}
+
\frac{ce\,\sigma_\theta^2\big(c+\beta_1^* e\lambda\big)}
{\big(c+e(\beta_1^*)^2\big)^2}
\right]>0.
\]
Finally, differentiating \eqref{we_clean} with respect to $e$ yields
\[
\frac{\partial w_e}{\partial e}
=
\frac{c\sigma_\theta^2(\lambda-\beta_1^*)}
{2\big(c+(\beta_1^*)^2e\big)^2}
\left[
c^2-(\beta_1^*)^3e^2(2\lambda-\beta_1^*)
\right].
\]
Thus, the sign of $\partial w_e/\partial e$ is the same as that of
$H(e)\equiv c^2-(\beta_1^*)^3e^2(2\lambda-\beta_1^*)$.
Since $H(0)=c^2>0$ and $H(.)$ is strictly decreasing in $e$
(because $e^2$, $(\beta_1^*)^3$, and $(2\lambda-\beta_1^*)$ are non-decreasing in $e$),
while $\lim_{e\to\infty}H(e)=c^2-e^2\lambda^4\to -\infty$,
there exists a unique threshold $\hat{e}>0$ at which $H(\hat{e})=0$.
For $e<\hat{e}$, $w_e$ is increasing; for $e>\hat{e}$, $w_e$ is decreasing.$\Box$
\medskip\\
\textbf{Proof of Lemma \ref{lem}:}
The sender's expected utility is
\begin{align*}
\mathbb{E}[u(r;s,x)]
&= \mathbb{E}[P(r)]
-\mathbb{E}\bigg[\frac{c}{2} (r-s-x)^2\bigg]
-\mathbb{E}\bigg[\frac{e}{2}\big(P(r)-\mathbb{E}(\theta| s)\big)^2\bigg]=-\frac{(\beta_1^*)^2}{2c}
-\frac{\sigma_\theta^2 e}{2}\left( \lambda-\beta_1^*\right).\Box
\end{align*}
\medskip\\
\textbf{Proof of Proposition \ref{P14}:}
Differentiating $W$ with respect to $c$ and using (\ref{sig}) yields
\begin{eqnarray}
\frac{\partial W}{\partial c}
&=&
\frac{\beta_1^{*2}+c\left(c e\sigma_\theta^2-2\beta_1^*\right)
\frac{\partial\beta_1^*}{\partial c}}
{2c^2}=\frac{\beta_1^{*2}}{2c^2}
\left(
1+
\frac{e(\beta_1^*-\lambda)(-2\beta_1^*+ce\sigma_\theta^2)}
{c+(\beta_1^*)^2 e}
\right),\nonumber \\ \label{wel22}
&=&
\frac{(\Delta\beta_1^*-1)^2}
{2e^2\big(\Delta(\beta_1^*)^2-2\beta_1^*+\lambda\big)}
\underbrace{\left[
\frac{\lambda(\lambda\Delta-1)}{(\lambda-\beta_1^*)^2}
-\Delta-e^2\sigma_\theta^2
\right]}_{L(c)}.
\end{eqnarray}
Note that
$q(\beta)\equiv \Delta\beta^2-2\beta+\lambda$ is increasing on
$[1/\Delta,\infty)$ with
$q(1/\Delta)=\lambda-1/\Delta>0$, so that the factoring term preceding the brackets is positive. Hence, $\frac{\partial W}{\partial c}\propto L(c)$. Because $\beta_1^*$ depends on $c$ only via $\beta_1^*$, which is decreasing in $c$, $L(c)$ is decreasing in $c$, it is readily verified that  $L(c)$ is also
strictly decreasing in $c$. To examine whether $L(c)$ changes signs, consider first $c\to 0$. Corollary \ref{e13} implies
$\beta_1^*\to\lambda$, which implies $L(c)\to \infty$. Next, as $c\to\infty$, the equilibrium converges to the pure lying-cost benchmark, so that
$\beta_1^*\to \frac1\Delta$.
Therefore:
\[
\lim_{c\to\infty}L(c)
=
\frac{\lambda(\lambda\Delta-1)}{\left(\lambda-\frac1\Delta\right)^2}
-\Delta-e^2\sigma_\theta^2
=\sigma_\theta^2\left(\frac{\Delta}{\lambda\sigma_x^2}-e^2\right)
.
\]
It then follows that, if $e\leq \hat{e}\equiv \frac{1}{\sigma_x}\sqrt{\frac{\Delta}{\lambda}}$, $W$ is decreasing in $c$ and, otherwise, it is inverse U-shaped in $c$.
As $c\to0$, 
$\beta_1^*\to\lambda$, which, from \eqref{trl1},
\[
W
=
-\frac{(\beta_1^*)^2}{2c}
-\frac{\sigma_\theta^2 e}{2}(\lambda-\beta_1^*)
\to -\infty.
\]

Similarly, differentiating $W$ with respect to $e$ gives \[ \frac{\partial W}{\partial e}
=
\frac12\left(
-\sigma_\theta^2\lambda
+
e\sigma_\theta^2\frac{\partial\beta_1^*}{\partial e}
+
\beta_1^*
\left(\sigma_\theta^2-\frac{2}{c}\frac{\partial\beta_1^*}{\partial e}\right)
\right)=\frac{(\beta_1^*-\lambda)(2\beta_1^{*3}+c^2\sigma_\theta^2)}
{2c(c+(\beta_1^*)^2 e)}<0,\]
using  \eqref{lkx} and \eqref{der} to insert \[
\frac{\partial\beta_1^*}{\partial e}
=
\frac{(\beta_1^*)^2(\lambda-\beta_1^*)}
{c+(\beta_1^*)^2 e}\] and simplify the last equality.$\Box$
\medskip\\
\textbf{Proof of Corollary \ref{JU}:} We know from (\ref{bias1})-(\ref{ie1}) that a fully-informative $\beta_1^*\rightarrow \lambda $ message is achieved if and only if $e\rightarrow \infty$, unless $c\rightarrow 0$ but this would imply unbounded sender disutility and thus is worse than in any $e\rightarrow \infty$ with finite sender expected utility.
Define $h=1/e$ and let us examine the limit of $W$ as $h\rightarrow 0$. Reorganizing the expression of the sender's welfare, 
\begin{eqnarray*}
W&=&-\frac{(\beta_1^*)^2}{2 c}+\frac{\sigma_\theta^2 }{2} \frac{\beta_1^*-\lambda}{h}\\
W_0\equiv\lim_{h \rightarrow 0} W &=&-\frac{1}{2c}\lambda^2 +\frac{\sigma_\theta^2 }{2}\frac{\partial {\beta_1^*}}{\partial h} |_{h\rightarrow 0}.
\end{eqnarray*}
The derivative in the second term is
$$\frac{\partial \beta_1^*}{\partial h}=\frac{c (\beta_1^* \Delta-1) }{\Gamma'(\beta_1^*)}\rightarrow -\frac{c \sigma_x^2}{\sigma_{\theta}^2}.$$
Reinjecting the above into $W_0$, the sender's welfare is strictly concave in $c$ and, taking the first-order condition in $c$, 
\begin{equation*}
\frac{\partial W_0}{\partial c}=\frac{\lambda^2}{c^2} - \sigma_x^2,
\end{equation*}
has a unique interior maximum $c^{**}_\infty=\frac{\lambda}{\sigma_x}$. Evaluating $W_0$ at this solution yields $W^{**}$ in the Proposition.$\Box$
\medskip\\
\textbf{Proof of Proposition \ref{PT_compare}:}
Let $e \rightarrow \infty$, so that the partition becomes arbitrarily fine and let $h_i\equiv t_{i+1}-t_i$. A second-order expansion of the truncated mean on cell \(i\) yields
\begin{equation}
\label{A2}
\mathbb E[\lambda s\mid t_i\le \lambda s<t_{i+1}]=
\frac{t_i+t_{i+1}}{2}+\underbrace{\frac{f'(\frac{t_i+t_{i+1}}{2})}{12f(\frac{t_i+t_{i+1}}{2})}\,h_i^2+o(h_i^2)}_{\delta_i}. \end{equation}

Recall  the cutoff indifference condition is $t_i+\frac{1}{e}=\frac{P(t_{i-1})+P(t_i)}{2}$, so that substituting the approximation in (\ref{A2}) yields
\begin{equation}
h_i-h_{i-1}
+
\frac{f'(t_i)}{6f(t_i)}\big(h_i^2+h_{i-1}^2\big)
=
\frac{4}{e}
+
o(h_i^2+h_{i-1}^2). \label{A3}
\end{equation}
This can be written in differential form:
\begin{equation}
\frac12\frac{d}{dt}\big(h(t)^2\big)
+\frac{f'(t)}{3f(t)}\,h(t)^2
=
\frac{4}{e}+o(e^{-1}).\label{A4a}
\end{equation}

Multiplying both sides by $f(t)^{2/3}$ and using the chain rule:
\[
\frac{d}{dt}\Big(h(t)^2 f(t)^{2/3}\Big)
=
\frac{8}{e}f(t)^{2/3}+o(e^{-1})f(t)^{2/3}.
\]
Integrating from $-\infty$ to $t$,
\begin{equation}\label{A4}
h(t)^2
=
\frac{8}{e}f(t)^{-2/3}\int_{-\infty}^t f(u)^{2/3}\,du
+o(e^{-1}).
\end{equation}
Next, consider the expected deception loss:
\[
w_e^P
\equiv
\frac{e}{2}\sum_i\int_{t_i}^{t_{i+1}}(P(t_i)-z)^2 f(z)\,dz.
\]
From a Taylor expansion, the integral can be written as
(proved below)\begin{equation}
\label{fik}
\int_{t_i}^{t_{i+1}}(P(t_i)-z)^2 f(z)\,dz
=
\frac{f\!\left(\frac{t_i+t_{i+1}}{2}\right)}{12}h_i^3+o(h_i^3).
\end{equation}
To prove (\ref{fik}), we can substitute in \eqref{A2} to show
\begin{eqnarray*}
\int_{t_i}^{t_{i+1}}(P(t_i)-z)^2 f(z)\,dz
&=&
\int_{t_i}^{t_{i+1}}(\frac{t_i+t_{i+1}}{2}+\delta_i-z)^2 f(z)\,dz\\
&=&\int_{-h_i/2}^{h_i/2}(\delta_i-y)^2 f(\frac{t_i+t_{i+1}}{2}+y)dy,\\
&=& 
f(\frac{t_i+t_{i+1}}{2})\int_{-h_i/2}^{h_i/2}(\delta_i-y)^2\,dy
+O(h_i)\int_{-h_i/2}^{h_i/2}(\delta_i-y)^2\,dy,
\end{eqnarray*}
where the second equality follows from a change of variables \(z=\frac{t_i+t_{i+1}}{2}+y\) and the third equality follows from the Taylor expansion $f(\frac{t_i+t_{i+1}}{2}+y)=f(\frac{t_i+t_{i+1}}{2})+O(h_i)$. Next,
\[
\int_{-h_i/2}^{h_i/2}(\delta_i-y)^2\,dy
=
\int_{-h_i/2}^{h_i/2}y^2\,dy+\delta_i^2 h_i=
\frac{h_i^3}{12}+O(h_i^5),
\]
which completes the proof of (\ref{fik}).
This equation can then be substituted into $w_e^P$ and noting that the resulting term is a Riemann sum, as the partition becomes arbitrarily fine, it simplifies to:
\begin{equation*}
w_e^P+o(1)
=
\frac{e}{24}\sum_i f\!\left(\frac{t_i+t_{i+1}}{2}\right)h_i^3=\frac{e}{24}\int f(t)\,h(t)^2\,dt.
\end{equation*}

Substituting \eqref{A4} into the above and using the Gaussian density:
\begingroup\footnotesize
\begin{eqnarray}
w_e^P+o(1)
&=&
\frac13\int
f(t)^{1/3}
\left(\int_{-\infty}^t f(u)^{2/3}\,du\right)dt
=\frac{1}{3\sqrt{2\pi\lambda\sigma_\theta^2}}
\int\int_{-\infty}^t
\exp\!\left(-\frac{t^2+2u^2}{6\lambda\sigma_\theta^2}\right)\,du\,dt \\
&=&\frac{1}{3\sqrt{4\pi\lambda\sigma_\theta^2}}
\frac12
\iint
\exp\!\left(-\frac{t^2+y^2}{6\lambda\sigma_\theta^2}\right)\,dy\,dt
=
\frac{\sigma_\theta}{2}\sqrt{\pi\lambda}.\Box
\end{eqnarray}
\endgroup
\medskip\\
\textbf{Proof of comparative statics in Proposition \ref{equi_genmix}:} The steps to derive the main proposition are almost identical to the baseline so, to save space, are delegated to a supplementary appendix. However, we prove next the comparative statics:
\begin{eqnarray*}
\frac{\partial \beta_1^*}{\partial e}&=&-\frac{\partial \Gamma^c(\beta_1^*)/\partial e}{(\Gamma^c)'(\beta_1^*)}\propto  \frac{\partial \Gamma^c(\beta_1^*)}{\partial e}= \beta_1^*(\lambda-\beta_1^*)>0 \\
\frac{\partial \beta_1^*}{\partial c}&\propto & \frac{\partial \Gamma^c(\beta_1^*)}{\partial c}= 1-\Delta\beta_1^*<0\\
\frac{\partial \beta_1^*}{\partial k}&\propto & \frac{\partial \Gamma^c(\beta_1^*)}{\partial k}= \frac{1}{2}\big(\lambda-\Delta(\beta_1^*)^2\big)\propto c-\frac{e\lambda}{\Delta} \propto \bar \beta_1^*-\beta_1^*.
\end{eqnarray*}
Thus, stronger complementarities move $\beta_1^*$ toward the pure complementarity benchmark $\bar\beta_1^*$. $\Box$
\medskip\\
\textbf{Proof of Corollary \ref{sigma_bias_gen}:}\ As shown below, the residual uncertainty only depends on $k$ via its effect on $\beta_1^*$:
\[
\Sigma
=
\sigma_\theta^2(1-\beta_1^*\alpha_1^*)
=
\sigma_\theta^2(1-\lambda)+\sigma_\theta^2(\lambda-\beta_1^*)\alpha_2^*=\sigma_\theta^2(1-\lambda)
+
\sigma_\theta^2\frac{(\lambda-\beta_1^*)^2}{\Delta(\beta_1^*)^2-2\beta_1^*+\lambda} .
\]
Differentiating with respect to $\beta_1^*$ yields
\[
\frac{d\Sigma}{d\beta_1^*}
=
-\frac{2\beta_1^*\,\sigma_\theta^2(\lambda-\beta_1^*)(\Delta\lambda-1)}
{\left(\Delta(\beta^*_1)^2-2\beta_1^*+\lambda\right)^2}
<0,
\]
which implies\[
\frac{\partial \Sigma}{\partial e}
=
\frac{d\Sigma}{d\beta_1^*}\frac{\partial \beta_1^*}{\partial e}<0,
\qquad
\frac{\partial \Sigma}{\partial c}
=
\frac{d\Sigma}{d\beta_1^*}\frac{\partial \beta_1^*}{\partial c}>0,\qquad \frac{\partial \Sigma}{\partial k}
=
\frac{d\Sigma}{d\beta_1^*}\frac{\partial \beta_1^*}{\partial k}\propto \beta_1^*-\bar\beta_1^*.
\]
For the bias, Proposition~\ref{equi_genmix} readily yields\[
 \quad B=-\frac{\beta_0^*}{\beta_1^*}=m_x+\frac{\beta_1^*}{c+\frac{k}{2}\beta_1^*},
\]
so that\[
\frac{\partial B}{\partial c}
=
\frac{c\,\frac{\partial \beta_1^*}{\partial c}-\beta_1^*}
{\left(c+\frac{k}{2}\beta_1^*\right)^2}<0,
\qquad
\frac{\partial B}{\partial e}
=
\frac{c\,\frac{\partial \beta_1^*}{\partial e}}
{\left(c+\frac{k}{2}\beta_1^*\right)^2}>0.
\]
Similarly,
\[
\frac{\partial B}{\partial k}
=
\frac{c\,\frac{\partial \beta_1^*}{\partial k}-\frac12(\beta^*_1)^2}
{\left(c+\frac{k}{2}\beta_1^*\right)^2}.
\]
Hence, if $\beta_1^*\ge \bar\beta_1^*$, then $\partial \beta_1^*/\partial k\le 0$ and therefore $\partial B/\partial k<0$ (otherwise the sign is ambiguous).$\Box$
\medskip
\\\textbf{Proof of Corollary~\ref{welfare_gen}:}
It is convenient to define $d_1 \equiv r-s-x$ and $d_2 \equiv P(r)-\lambda s$, so that the sender's expected utility is $W
=
\mathbb{E}[P(r)]
-\frac{c}{2}\mathbb{E}[d_1^2]
-\frac{e}{2}\mathbb{E}[d_2^2]
-\frac{k}{2}\mathbb{E}[d_1 d_2]$. Rewriting the sender's first-order condition (equation (B.10) in the online appendix) with this notation:
\[
d_1=\frac{\beta_1^*}{ c+\frac{k}{2}\beta_1^*}-\frac{e\beta_1^*+\frac{k}{2}}{ c+\frac{k}{2}\beta_1^*}\,d_2.
\]
Taking expectations, $\mathbb{E}[d_1]=\beta_1^*/(c+\frac{k}{2}\beta_1^*)$.
Moreover,
\[
\mathbb{E}[d_1^2]
=
\frac{(\beta^*_1)^2}{\Big( c+\frac{k}{2}\beta_1^*\Big)^2}
+\frac{\big(e\beta_1^*+\frac{k}{2}\big)^2}{\Big( c+\frac{k}{2}\beta_1^*\Big)^2}\,\mathbb{E}[d_2^2],
\qquad
\mathbb{E}[d_1 d_2]
=
-\frac{e\beta_1^*+\frac{k}{2}}{c+\frac{k}{2}\beta_1^*}\,\mathbb{E}[d_2^2].
\]
Finally, $\mathbb{E}[d_2^2]$ is the variance of the difference between the posterior based on $s$ and the posterior
based on $r$. By the law of total variance, $
\mathbb{E}[d_2^2]
=Var(\theta| r)-Var(\theta| s).
$
Using Corollary~\ref{sigma_bias_gen},
$
\mathbb{E}[d_2^2]
=
\sigma_\theta^2(\lambda-\beta_1^*)\alpha_2^*
=
\sigma_\theta^2(\lambda-\beta_1^*)\frac{c+\frac{k}{2}\beta_1^*}{c+k\beta_1^*+e(\beta_1^*)^2}$.   Substituting these expressions into $W$
and simplifying yields
\[
W
=
-\frac{c\,(\beta^*_1)^2}{2( c+\frac{k}{2}\beta_1^*)^2}
-\frac{1}{2}\mathbb{E}[d_2^2]\Big(\frac{c(e\beta_1^*+\frac{k}{2})^2}{( c+\frac{k}{2}\beta_1^*)^2}+e-\frac{k(e\beta_1^*+\frac{k}{2})}{c+\frac{k}{2}\beta_1^*}\Big).
\]
The bracket equals $\frac{4ce-k^2}{4}\cdot \frac{c+k\beta_1^*+e(\beta_1^*)^2}{( c+\frac{k}{2}\beta_1^*)^2}$: using the above expression for
$\mathbb{E}[d_2^2]$ yields  \eqref{eq:W_gen}.$\Box$
\medskip\\
\textbf{Proof of Corollary~\ref{fullinfo_gen}:}
Assume \ $k^2\le 4ce$. Then the quadratic form $
\frac{c}{2}d_1^2+\frac{e}{2}d_2^2+\frac{k}{2}d_1d_2$
is nonnegative for all $(d_1,d_2)$, so the sender's payoff satisfies $u(r;s,x)\le P(r)$ pointwise. Taking
expectations and using $\mathbb{E}[P(r)]=\mathbb{E}[\theta]=0$ yields $W\le 0$ for any policy, hence
$\sup W_\infty\le 0$.
To show attainability, fix any $c>0$ and consider a sequence with $e_i\to\infty$ and $k_i=2\sqrt{c e_i}$.
Let $\beta^{*}_{1,i}$ denote the corresponding equilibrium sensitivity, i.e.\ the positive root of \eqref{eq:Gamma_gen_restate}.
Rewrite \eqref{eq:Gamma_gen_restate} as $-(e_i+\sqrt{c e_i}\,\Delta)(\beta^*_{1,i})^2+(e_i\lambda-c\Delta)\beta^{*}_{1,i}+c+\sqrt{c e_i}\,\lambda=0.$
Divide by $e_i$ and let $e_i\to\infty$. Since $\sqrt{c/e_i}\to 0$, the limiting equation is
$-(\beta_{1,i}^*)^2+\lambda\beta_{1,i}^*=0$, so $\beta^{*}_{1,i}\to\lambda$.
Next, the language noise pass-through coefficient satisfies
\[
\alpha_{2,i}^*
=\frac{c+\frac{k_i}{2}\beta^{*}_{1,i}}{c+k_i\beta^{*}_{1,i}+e_i(\beta^*_{1,i})^2}
=
\frac{c+\sqrt{c e_i}\,\beta^{*}_{1,i}}{c+2\sqrt{c e_i}\,\beta^{*}_{1,i}+e_i(\beta^{*}_{1,i})^2}
\to 0,
\]
so $\Sigma_i-Var(\theta| s)=\sigma_\theta^2(\lambda-\beta^{*}_{1,i})\alpha_{2,i}^*\to 0$ and therefore $\Sigma_i\to Var(\theta| s)$.
Under $k_i^2=4c e_i$, the second term in \eqref{eq:W_gen} is zero,
and thus
\[
W_i
=
-\frac{c\,(\beta^*_{1,i})^2}{2\big(c+\frac{k_i}{2}\beta^{*}_{1,i}\big)^2}
=
-\frac{c\,(\beta^*_{1,i})^2}{2\big(c+\sqrt{c e_i}\,\beta^*_{1,i}\big)^2}
\to 0.
\]
Combined with $\sup_{\omega \in \Omega} W_\infty(\omega)\le 0$, this proves
$\sup_{\omega \in \Omega} W_\infty(\omega) = 0$. 
$\Box$

\end{singlespace}

\section*{Figure Legends}

\clearpage
\setcounter{equation}{0}
\setcounter{proposition}{0}
\setcounter{corollary}{0}
\setcounter{definition}{0}
\setcounter{figure}{0}
\setcounter{footnote}{0}

\renewcommand{\theequation}{B.\arabic{equation}}
\renewcommand{\theproposition}{B.\arabic{proposition}}
\renewcommand{\thecorollary}{B.\arabic{corollary}}
\renewcommand{\thedefinition}{B.\arabic{definition}}
\renewcommand{\thefigure}{B.\arabic{figure}}

\section*{{\sc Truth and Deception}\\ \vspace{0.3cm} {\large Supplementary (Online) Appendix}}
\thispagestyle{fancy}

\vspace{0.5cm}

\begin{singlespace}

\subsection*{B.1. Equilibrium with Negative Belief Sensitivity}

\paragraph{}
So far, we restricted attention to equilibria with $\beta_1^*\ge 0$, in which higher messages are issued following better news. However, the presence of deception costs allows for a qualitatively different type of self-fulfilling equilibrium. Suppose the receiver conjectures that higher messages reflect \textit{worse} news, i.e., $\hat{\beta}_1<0$. If the sender were to follow a strategy with $\beta_1>0$, the receiver would systematically misinterpret reports, generating large deception costs. Anticipating this, the sender may instead optimally choose a reporting strategy with $\beta_1^*<0$, thereby validating the receiver’s pessimistic beliefs.\footnote{See also \cite{stozha23} for a model in which higher-order uncertainty causes a more positive signal to be interpreted negatively. While the mechanism is related (beliefs about how messages are interpreted matter for equilibrium), their focus is on global games rather than rational expectations communication.} 

It may appear that these equilibria require deception costs $e$ to be large relative to lying costs $c$, since strategies with $\beta_1<0$ can generate substantial lying costs. However, as we show below, a negative-sensitivity equilibrium exists for any $e>0$.

\begin{proposition}\label{jui}
If $e,c>0$, there exists a unique equilibrium with $\underline{\beta}_1^*<0$, where
$\underline{\beta}_1^*$ is the negative root of equation (7) in the main text. As $c\to0$,$\underline{\beta}_1^*\to0$,
so the equilibrium converges to uninformative babbling.
\end{proposition}

This equilibrium connects  to babbling equilibria in the cheap-talk limit. As $c\to 0$, the positive root $\beta_1^*$ does not converge to any cheap-talk equilibrium, because the equilibrium becomes fully revealing in the limit (Corollary 2 in the main text), whereas all cheap-talk equilibria with $c=0$ necessarily involve an informational loss \citep{crasob82}. By contrast, the negative root converges to babbling, $\underline{\beta}_1^*\to 0$, which is the only linear cheap-talk equilibrium and therefore the natural continuous extension of the equilibrium set when lying costs vanish.

\begin{figure}
\begin{center}\includegraphics[scale=.55]{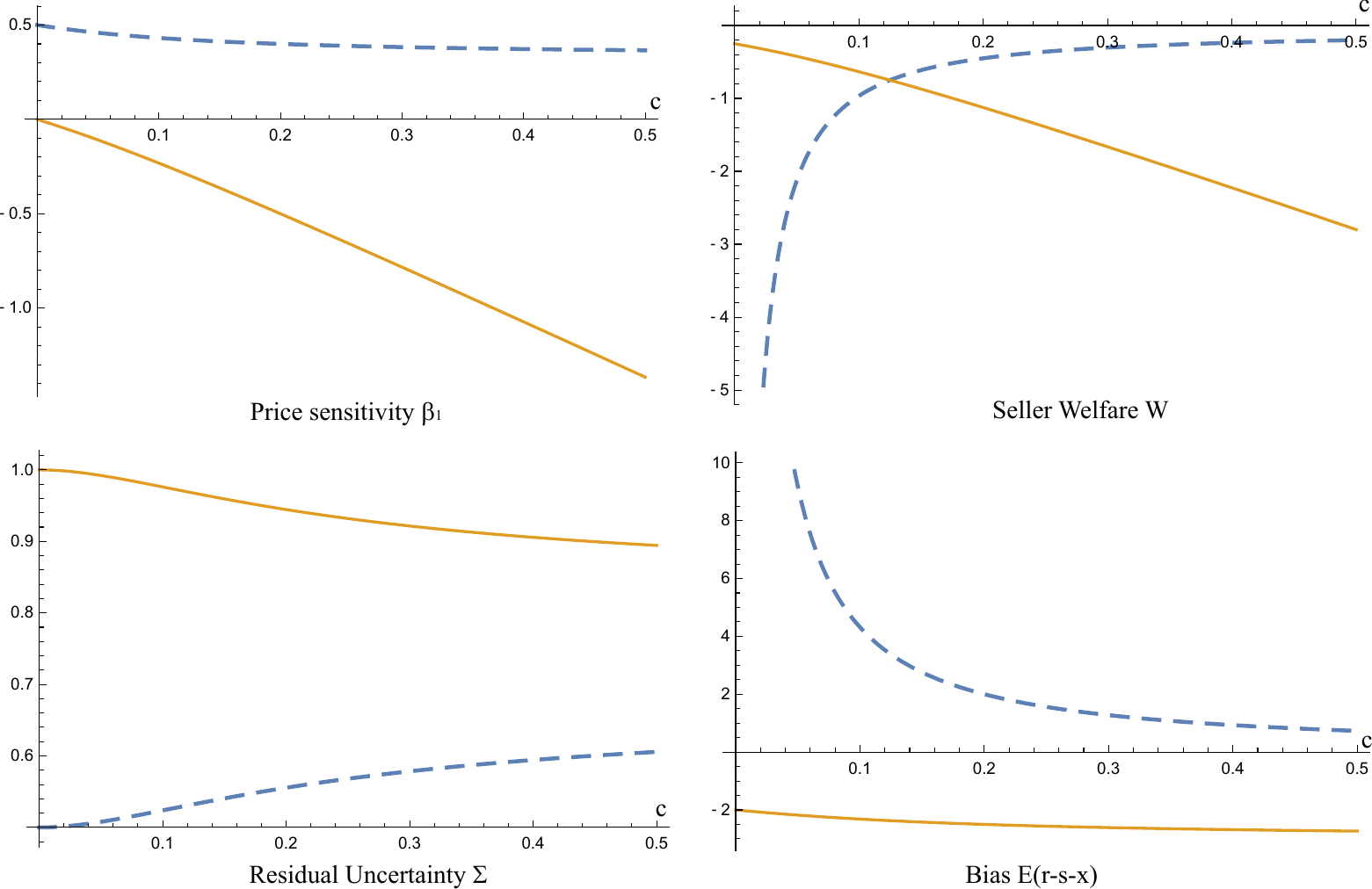}
\caption{Comparative Statics with $\underline{\beta}_{1}^*<0$ (solid) versus $\beta_{1}^*$ (dashed), with $\sigma_\theta,\sigma_\epsilon,\sigma_x,e=1$. \label{X01}}

\vspace{0.3cm}
Alt text: Four line graphs plotting the negative-sensitivity equilibrium against the lying cost. Top left: price sensitivity, with the solid curve falling from zero and the dashed curve near one half. Top right: seller welfare, with the solid curve decreasing and the dashed curve rising from minus five toward zero. Bottom left: residual uncertainty, with the solid curve near one and the dashed curve rising from one half. Bottom right: bias, with the dashed curve decreasing from ten toward one and the solid curve near minus two.\end{center}
\end{figure}

In addition, an equilibrium with $\underline{\beta}_1^*<0$ may be preferred by the sender to the positive root. Although the sender is strictly worse off under $\underline{\beta}_1^*$ than under $\beta_1^*$ when $c$ is large, the reverse holds when $c$ is small. As shown in the top-right panel of Figure~\ref{X01}, under the positive root $\beta_1^*$ the sender incurs unbounded lying costs as $c\to 0$, since equilibrium bias explodes. By contrast, under $\underline{\beta}_1^*=0$ the report is uninformative, bias is zero, and deception penalties remain bounded at $\frac{e}{2}\big(Var(\theta)-Var(\theta|s)\big)$. Hence, when lying costs are small, the sender may prefer to coordinate on the negative-sensitivity equilibrium, effectively selecting a less informative but less costly communication regime.

The informational implications are illustrated in the bottom-left panel of Figure~\ref{X01}. When $c$ is small, $\underline{\beta}_1^*\approx 0$ and residual uncertainty is maximal: the report is effectively uninformative. As $c$ increases, the magnitude $|\underline{\beta}_1^*|$ grows, increasing the sensitivity of the report to private information. In the limit $c\to\infty$, residual uncertainty converges to
\[\Sigma
\longrightarrow
\sigma_\theta^2\!\left(1-\lambda+\frac{1}{\Delta}\right)> \mathrm{Var}(\theta| s),\]
as stated in Proposition~\ref{jui}. Unlike the positive root, the negative-sensitivity equilibrium does not achieve full revelation even in the limit $c\to\infty$: residual uncertainty remains strictly above $\mathrm{Var}(\theta|s)$ because the sender’s reversed reporting strategy conveys less information per unit of lying cost. Thus, while the positive root achieves near-perfect communication with vanishing lying costs, the negative root converges to a noisier limit regardless of how large $c$ becomes. In both cases, the sender’s expected penalties become very large.

\subsection*{B.2. Uncertain benefits}
\paragraph{}
In our baseline model, uncertainty is composed of the fundamental uncertainty about the state $s$ and the language noise $x$, but the misalignment with the receiver is common-knowledge. We extend the model to plausible environments where the sender may not know the extent of the misalignment. In practice, for example, an agent's compensation structure, reputational concerns, or general propensity to issue biased advice may not be fully observable. To capture this,  \citet{fisver00} consider a model in which the sender's utility is $yP(r)$, where $y$ is an idiosyncratic
preference shock privately observed by the sender. Adapting their assumption to our model, we consider a generalized version of the preference:
\begin{equation}\label{FV}
u(r;s,x,y)
= y P(r)
- \frac{c}{2}\,(r-s-x)^2
- \frac{e}{2}\Big(P(r)-\mathbb{E}(\theta| s)\Big)^2,
\end{equation}
where 
$y\sim N(m_y,\sigma_y^2)$ is independent of all other shocks.

Note that it is critical for our model to retain the possibility of language noise since, otherwise, a regulation could enable forcing truthful reporting by applying an arbitrarily large penalty $c$. Hence, we retain $\sigma_x>0$ in this extension. The formal analysis is identical to the baseline model, so we omit it to save space.
\begin{proposition}\label{prop_FVy_equil}
For any $c>0$, there exists a unique LREE with positive belief sensitivity $\beta_1^*>0$ with:
\begin{itemize}

\item[(i)]  The receiver's belief is $P^*(r)=\beta_0^*+\beta_1^* r$, where $\beta_1^*\in (0,\lambda)$ is the unique root of
\begin{equation}
\label{GammaFVy}
\begin{aligned}
\Gamma^{f}(\beta_1)
\equiv\;
\frac{\sigma_\theta^2}{\lambda}\sigma_y^2\,\beta_1^3
+ ce\,\frac{\sigma_\theta^4}{\lambda}\,\beta_1^2 + c\sigma_\theta^4\!\left(\frac{c\Delta}{\lambda}-e\right)\beta_1
- c^2\frac{\sigma_\theta^4}{\lambda}
=0,
\end{aligned}
\end{equation}
and the intercept is
\begin{equation}\label{beta0FVy}
\beta_0^{*}
=
-\beta_1^{*}\left(m_x+\frac{\beta_1^{*}}{c}m_y\right).
\end{equation}

\item[(ii)] The reporting strategy is linear,
$R^*(s,x,y)=\alpha_0^{*}+\alpha_1^{*}s+\alpha_2^{*}x+\alpha_3^{*}y$, with
\[
\alpha_0^{*}
=
-\frac{\beta_0^{*}\beta_1^{*}e}{c+(\beta_1^{*})^2e},\qquad
\alpha_1^{*}
=
\frac{c+\lambda\beta_1^{*}e}
{c+(\beta_1^{*})^2e},
\qquad
\alpha_2^{*}
=
\frac{c}{c+(\beta_1^{*})^2 e},
\qquad
\alpha_3^{*}
=
\frac{\beta_1^{*}}{c+(\beta_1^{*})^2 e}.
\]
\end{itemize}
Further, $\beta_1^{*}$ is increasing in $e$ and decreasing in $\sigma_y^2$ (the effect of the lying cost $c$ is generally ambiguous).

\end{proposition}

To summarize, the reporting strategy adds another source of uncertainty tied to the sender's uncertain benefit, and this source becomes more important when the price response is stronger. This has a critical effect on the analysis because, unlike in the baseline model, higher belief sensitivity now causes more motive noise to enter the message, which further muddles the signal. Consistent with this intuition, greater benefit noise reduces belief sensitivity. While deception costs still discipline the sender, the effect of the lying cost is now more ambiguous: although higher $c$ still induces more language noise to enter the message, it also shifts the reporting strategy away from noisy motive shocks and toward the language benchmark, which can improve information transmission.

The expected bias retains a simple expression, generalizing the unit benefit in the baseline to the mean benefit $m_y$:  
\[
B= m_x+\frac{\beta_1^{*}}{c}m_y,
\]
and implying that the comparative statics in the deception cost $e$, which only affect bias via $\beta_1^*$, are unchanged. By contrast, the effect of the lying cost becomes more ambiguous in situations where $\beta_1^*$ increases in $c$. Intuitively, if lying cost allow for enough filtering of the noisy benefit, it can increase $\beta_1^*$ sufficiently to (counter-intuitively) increase total bias. However, we show in the next Corollary that the increase in belief sensitivity is never sufficiently high to increase expected bias.

\begin{corollary}\label{P31_FVy}
 If $m_y>0$, the average bias $B$ is increasing in $e$, increasing in $m_x$ and $m_y$, and decreasing in $c$ and $\sigma_y^2$.
\end{corollary}

This is not the case for the residual uncertainty:
\begin{equation}\label{varFVy}
\Sigma =\sigma_\theta^2\Big(1-\beta_1^{*}\alpha_1^{*}\Big)
=\sigma_\theta^2\left(1-\frac{\beta_1^{*}\big(c+e\beta_1^{*}\lambda\big)}{c+e(\beta_1^{*})^2}\right),
\end{equation}
which, by an intuition analogous to that driving $\beta_1^*$, is generally non-monotonic in the lying cost. This effect differs from \citet{fisver00} and is driven by the interaction between the language-noise channel and the uncertain benefit. By contrast, when $\sigma_x^2=0$, increasing $c$ improves precision because it reduces the pass-through of the unobservable motive shock $y$ without introducing any additional exogenous noise into the report.

\begin{corollary}\label{P31_FVy2}
The residual uncertainty $\Sigma$
is strictly decreasing in $e$ and increasing in $\sigma_y^2$. The effect of the lying cost $c$ on
$\Sigma$ is generally ambiguous, decreasing (increasing) in $c$ if $c$ is low (high).
\end{corollary}

Finally, we examine the sender's preference, noting that (as shown in prior literature) the sender now benefits from controlling the receivers' belief, by correlating posterior beliefs and $y$. This creates a channel, absent in the baseline model, through which persuasion creates ex-ante value $\mathbb{E}(y P(r))$ relative to a fully-informative signal $R(s,x,y)=s$. Because both $c$ and $e$ reduce the sender's control over beliefs, the sender has, in this model, stronger preferences for low enforcement and \cite{fisver00} show that there may be cases where the sender prefers to bear higher costs and have more discretion over the reporting process. 

\begin{corollary}\label{wel2_FVy}
Let $W^{pool}$ denote the sender welfare in the LREE with $c=0$, where the message is uninformative. Then
\begin{equation}
W-W^{pool}
=
\frac{\beta_1^*}{2}\left(
e\sigma_\theta^2+\frac{\beta_1^*}{c}\big(\sigma_y^2-m_y^2\big)
\right).
\end{equation}
Consequently, $W$ is decreasing in $|m_y|$. If $m_y=0$, then $W>0$ if and only if $\beta_1^{*}<1/\Delta$, or equivalently $\sigma_y^2>ce\,\sigma_\theta^2\Delta(\Delta\lambda-1)$. More generally, if $|m_y|\le \sigma_y$, then $W>W^{pool}$.
\end{corollary}

Corollary \ref{wel2_FVy} establishes that uncertain benefits create an additional reason for the sender to value discretion, since reporting can now correlate posterior beliefs with the private motive shock. This force weakens, but does not eliminate, the benefit of lying costs. In particular, if $|m_y|\le \sigma_y$, then the sender strictly prefers any informative LREE with $c>0$ to pooling. The receiver does so a fortiori, since pooling maximizes residual uncertainty. More generally, however, sender welfare need not be monotone in $c$: when the mean motive is sufficiently large relative to the uncertainty in motives, the sender may prefer pooling to equilibria with small lying costs.

In our final result, we explore how fully-informative communication may be achieved. Recall that, in the baseline, this yields an interior level of lying cost. We show, similarly, that the model requires a lying cost that is interior and can be uniquely characterized. 

\begin{corollary}\label{JU_FVy}
Assume $m_y>0$. Consider the set $\Omega$ of all sequences of penalty pairs $\omega=(c_i,e_i)_{i=1}^{\infty}$ such that, as $i\to\infty$, the associated residual uncertainty satisfies $\Sigma_i\to \mathrm{Var}(\theta| s)$ and the sender's welfare converges to $W_i \rightarrow W_\infty(\omega)$. Then $\sup_{\omega \in \Omega} W_\infty(\omega) = -\lambda m_y \sigma_x$, and this supremum can be approximated arbitrarily closely by sequences with $e_i\to \infty$ and $c_i\to c_\infty^{**}\equiv \lambda m_y/\sigma_x$.
\end{corollary}

Corollary \ref{JU_FVy} shows that the uncertain-benefit extension preserves the main design logic of the baseline model. To approach full revelation, the regulator must send the deception cost to infinity, since otherwise the private motive shock $y$ continues to contaminate the report. But this does not require the lying cost to become arbitrarily large. Once $e$ is sufficiently high, the asymptotic trade-off is again between two familiar forces: a higher $c$ reduces the average bias generated by the mean motive $m_y$, while a lower $c$ relaxes the sender's attachment to the noisy language benchmark. In the limit, the variance of the motive shock $\sigma_y^2$ affects how fast deception costs must grow, but it does not affect the optimal limiting lying cost itself. Thus, the optimal full-information design again features an interior amount of tolerance for lying, with the baseline formula modified by replacing the unit benefit with its mean $m_y$.

\subsection*{B.3. Deception penalty without knowledge of the state}

\paragraph{}

In the baseline model, we assumed that the deception penalty depends on the deviation between the receiver's posterior belief $P(r)$ and the sender's own expectation $\mathbb{E}[\theta| s]$. This is appropriate when the sender knows the state and internalizes the consequences of misleading the receiver, or when an enforcer can eventually observe an informative realization of the state and assess the damage caused by deception. In some settings, however, the underlying state may never be fully revealed. For example, courts may have difficulty disentangling the effect of a particular misstatement from other economic shocks. In such cases, deception may be assessed only based on observable reports and the language used, without direct access to the state.

To capture this extreme case, consider replacing the deception term by a penalty based on the posterior given the report and the language benchmark:
\begin{equation}\label{U11b}
u(r;s,x)=P(r)-\frac{c}{2} (r-s-x)^2
-\frac{e}{2}\big(P(r)-\mathbb{E}(\theta| s+x,r)\big)^2.
\end{equation}
Here, the last term penalizes deviations between the posterior $P(r)$ and the posterior that would be formed if the enforcer could use both $r$ and the language benchmark $s+x$, but not the private signal $s$ itself.

It may seem that this change is innocuous, since on the equilibrium path $(s+x,r)$ is a sufficient statistic for $s$ whenever the reporting strategy is linear with $\alpha_1\neq \alpha_2$. However, off the equilibrium path the sender can manipulate $\mathbb{E}(\theta| s+x,r)$ by choosing reports that distort the joint distribution of $(s+x,r)$, and this changes the incentives.

A key observation is that, in any linear equilibrium of this modified game, the sender's payoff depends on $(s,x)$ only through their sum $s+x$. This implies that any equilibrium must have $\alpha_1=\alpha_2$.

\begin{proposition}\label{LMB}
Any linear rational expectations equilibrium must satisfy $\alpha_1=\alpha_2$, so that $R(s,x)$ depends only on $s+x$. \end{proposition}

Intuitively, when the enforcer does not observe the underlying state, she cannot distinguish between bias driven by the sender's strategic misreporting and bias due to the language noise $x$. The sender can always choose a reporting strategy that leaves the conditional distribution of $(s+x,r)$ unchanged while manipulating beliefs, so any equilibrium that attempts to penalize deception based only on $(s+x,r)$ must have zero effective deception costs. In this extreme case the only constraint on communication comes from the lying cost, and deception penalties no longer improve information transmission.

\subsection*{B.4. Global Stability under Iterated Best Responses}
\label{app:iter}

\paragraph{}
To complement the fixed-point characterization, we show that the LREE is the unique globally stable fixed point of the best-response operator on the set of linear conjectures with non-negative slope. Start from any initial conjecture
\[
P_0(r)=\beta_{0,0}+\beta_{1,0}r,
\qquad
\beta_{0,0}\in\mathbb{R},
\quad
\beta_{1,0}\ge 0.
\]
Given $P_n(r)=\beta_{0,n}+\beta_{1,n}r$, let $R_n=\mathrm{BR}(P_n)$ denote the sender's best response:
\[
R_n(s,x)=\alpha_{0,n}+\alpha_{1,n}s+\alpha_{2,n}x,
\]
with
\begin{equation}\label{eq:alpha_iter}
\alpha_{0,n}=\frac{\beta_{1,n}}{c+e\beta_{1,n}^2}\big(1-e\beta_{0,n}\big),\qquad
\alpha_{1,n}=\frac{c+\lambda e\beta_{1,n}}{c+e\beta_{1,n}^2},\qquad
\alpha_{2,n}=\frac{c}{c+e\beta_{1,n}^2}.
\end{equation}
Given $R_n$, define the next conjecture by Bayesian updating:
\[
P_{n+1}(r)=\mathbb{E}[\theta| R_n(s,x)=r].
\]
Since $(\theta,s,x)$ are jointly Gaussian and $R_n$ is linear, $P_{n+1}$ is linear:
\begin{equation}\label{eq:beta_update_general}
\beta_{1,n+1}
=
\frac{\alpha_{1,n}\sigma_\theta^2}
{\alpha_{1,n}^2(\sigma_\theta^2+\sigma_\epsilon^2)+\sigma_x^2\alpha_{2,n}^2},
\qquad
\beta_{0,n+1}
=
-\beta_{1,n+1}\big(\alpha_{0,n}+\alpha_{2,n}m_x\big).
\end{equation}
Therefore, the evolution of $\beta_{1,n}$ depends only on $\beta_{1,n}$ itself, not on the intercept.

\begin{proposition}\label{prop:iter_convergence}
Assume $c>0$. For any initial linear conjecture $P_0(r)=\beta_{0,0}+\beta_{1,0}r$ with $\beta_{1,0}\ge 0$, the sequence $(P_n,R_n)_{n\ge 0}$ generated above satisfies:

\begin{enumerate}
\item[(i)] For every $n$, $P_n$ and $R_n$ are linear, with coefficients satisfying \eqref{eq:alpha_iter}--\eqref{eq:beta_update_general}.

\item[(ii)] The slope sequence satisfies the one-dimensional recursion
\begin{equation}\label{eq:Tdef_global}
\beta_{1,n+1}=\mathcal{T}(\beta_{1,n}),
\qquad
\mathcal{T}(\beta)
=
\frac{\lambda(c+e\lambda\beta)(c+e\beta^2)}
{(c+e\lambda\beta)^2+c^2(\lambda\Delta-1)}.
\end{equation}
Moreover,
\[
\beta_{1,n}\uparrow \beta_1^*
\quad\text{if }\beta_{1,0}<\beta_1^*,
\qquad
\beta_{1,n}\downarrow \beta_1^*
\quad\text{if }\beta_{1,0}>\beta_1^*,
\qquad
\beta_{1,n}=\beta_1^*
\quad\text{if }\beta_{1,0}=\beta_1^*,
\]
where $\beta_1^*$ is the unique positive root of $\Gamma(\beta_1)=0$ in equation (7) in the main text. In particular, $\beta_{1,n}\to\beta_1^*$.

\item[(iii)] The intercepts converge:
\[
\beta_{0,n}\to\beta_0^*
\qquad\text{with}\qquad
\beta_0^*=-\beta_1^*\Big(\frac{\beta_1^*}{c}+m_x\Big),
\]
and therefore
\[
(\alpha_{0,n},\alpha_{1,n},\alpha_{2,n})\to(\alpha_0^*,\alpha_1^*,\alpha_2^*).
\]
Hence $(P_n,R_n)\to(P^*,R^*)$, the LREE characterized in Proposition 1 in the main text.
\end{enumerate}
\end{proposition}

Note that the naive prior $P_0(r)=r$ and the skeptical prior $P_0(r)=0$ correspond to $(\beta_{0,0},\beta_{1,0})=(0,1)$ and $(0,0)$, respectively, and are therefore special cases of Proposition~\ref{prop:iter_convergence}.

\subsection*{B.5. Uniqueness of Equilibrium with Lying Costs}

\paragraph{} 
We now ask whether, once lying costs are strictly positive, there may exist equilibria outside the linear class that combine smooth and partitional report regions and improve upon the LREE. The next results show that, within the admissible regular class defined below, this is impossible. Although the definition allows partitional regions a priori, such regions cannot be sustained in equilibrium when $c>0$ and $e>0$. It follows that any admissible regular equilibrium must in fact be fully continuous; the regularity condition then implies that it is affine, and the earlier characterization of affine equilibria therefore implies that it is the LREE.
In what follows, we assume throughout that $c>0$ and $e>0$.

\begin{definition}\label{def:B5_admissible}
An equilibrium $(R,P)$ is \emph{admissible} if the report space $\mathbb{R}$ can be partitioned into finitely many closed intervals such that, on each interval, one of the following holds:
\begin{enumerate}
\item[(i)] $P$ is strictly increasing and $C^2$ on the interior of the interval; or
\item[(ii)] only finitely many reports in the interval are used on path, while $P$ is continuous on the whole interval and, on each open off-path gap, coincides with a polynomial of degree bounded uniformly across gaps.
\end{enumerate}
\end{definition}

We further impose the following regularity condition.

\begin{definition}\label{def:B5_regular}
An admissible equilibrium $(R,P)$ is \emph{regular} if:
\begin{enumerate}
\item[(i)] $P:\mathbb{R}\to\mathbb{R}$ is continuous and strictly increasing, and after refining the partition in Definition \ref{def:B5_admissible} if necessary, $P$ is $C^2$ on the interior of each interval with nonempty interior;
\item[(ii)] for every $(s,x)$, the sender's optimal report $R(s,x)$ is unique and interior; and
\item[(iii)] whenever $R(s,x)$ lies in the interior of a smooth interval, the second-order condition is strict:
\[
P''(R(s,x))
-c
-e\Big((P'(R(s,x)))^2+\big(P(R(s,x))-\lambda s\big)P''(R(s,x))\Big)<0.
\]
\end{enumerate}
\end{definition}

The proof proceeds in three steps. First, we show that when $c>0$ and $e>0$, every available report is the unique optimum for some sender type. Hence no open off-path gap can be sustained, and genuine partitional regions are impossible. Second, on each smooth region, the sender's first-order condition implies that for any fixed report the locus of types for which that report is stationary is a straight line in the $(s,x)$-plane. Third, uniqueness of the sender's optimal report together with the strict second-order condition rules out crossings. They must therefore be parallel, which forces the slope of $P(\cdot)$ to be constant and hence implies that the equilibrium is affine.\footnote{One may reasonably consider whether the LREE with negative belief sensivity $\underline{\beta}_1^*<0$, examined in B.1, is ruled out by this argument. It is not. However, this negative root never converges to fully-revealing belief and, in fact, (counter-intuitively) converges to babbling when $e$ becomes large.}

\begin{proposition}\label{prop:B5_onpath}
In any admissible equilibrium, every report $r\in\mathbb{R}$ is sent on path by some type $(s,x)$. Consequently, an admissible equilibrium cannot contain a partitional interval. Every admissible equilibrium is therefore continuous.
\end{proposition}

Proposition ~\ref{prop:B5_onpath} already rules out the possibility that, once $c>0$, one can sustain a hybrid equilibrium by keeping open intervals of reports off path and filling them with a smooth continuation of beliefs. The only remaining candidate equilibria are continuous.

\begin{proposition}\label{prop:B5_unique}
Any admissible equilibrium that is regular coincides with the unique LREE characterized in Proposition 1 in the main text. In particular,
\[
P(r)=\beta_0^*+\beta_1^*r
\qquad\text{and}\qquad
R(s,x)=\alpha_0^*+\alpha_1^*s+\alpha_2^*x,
\]
where $(\beta_0^*,\beta_1^*,\alpha_0^*,\alpha_1^*,\alpha_2^*)$ are the coefficients in Proposition 1 in the main text.
\end{proposition}

\subsection*{B.6. Supplementary Appendix Proofs}

\textbf{Proof of Proposition 5 in the main text:} Taking the derivative of equation (20) in the main text with respect to $r$ yields the first-order condition
\begin{equation}\label{LX_gen}
\beta_1
- c\big(R(s,x)-s-x\big)
- e\beta_1\big(P(R(s,x))-\mathbb{E}(\theta| s)\big)
-\frac{k}{2}\Big(\big(P(R(s,x))-\mathbb{E}(\theta| s)\big)+\beta_1\big(R(s,x)-s-x\big)\Big)=0.
\end{equation}
The second-order condition is $\frac{\partial^2 u}{\partial r^2} = -(c+k\beta_1+e\beta_1^2)$,
which is negative for $\beta_1>0$.
Solving the above equation for $R(s,x)$, we obtain the linear best response
\begin{equation}\label{N11_gen}
R(s,x)=\underbrace{\frac{\beta_1-(e\beta_1+\frac{k}{2})\beta_0}{c+k\beta_1+e\beta_1^2}}_{\equiv\alpha_0}
+\underbrace{\frac{c+\frac{k}{2}\beta_1+\lambda\left(e\beta_1+\frac{k}{2}\right)}{c+k\beta_1+e\beta_1^2}}_{\equiv\alpha_1}s
+\underbrace{\frac{c+\frac{k}{2}\beta_1}{c+k\beta_1+e\beta_1^2}}_{\equiv\alpha_2}x.
\end{equation}
To impose rational expectations, let write the report as $ 
r=\alpha_0+\alpha_1 s+\alpha_2 x.$ Rearranging \eqref{N11_gen} returns:
\begin{equation}\label{signalform_gen}
\frac{r}{\alpha_1}-\frac{\alpha_0}{\alpha_1}-\frac{\alpha_2}{\alpha_1}m_x
= \theta+\epsilon+\frac{\alpha_2}{\alpha_1}(x-m_x),
\end{equation}so that the receiver's posterior expectation is
\begin{align}
P(r)
&=\mathbb{E}(\theta| r)=
\frac{\sigma_\theta^2}{\sigma_\theta^2+\sigma_\epsilon^2+\frac{\alpha_2^2}{\alpha_1^2}\sigma_x^2}
\cdot
\frac{r-\alpha_0-\alpha_2 m_x}{\alpha_1}\nonumber\\
&=
\frac{1}{\frac{\alpha_1}{\lambda}+\frac{\alpha_2^2}{\alpha_1}\left(\Delta-\frac{1}{\lambda}\right)}
\bigl(r-\alpha_0-\alpha_2 m_x\bigr)=\underbrace{-\beta_1(\alpha_0+\alpha_2 m_x)}_{=\beta_0}
+\underbrace{\frac{\alpha_1}{\frac{\alpha_1^2}{\lambda}+\alpha_2^2\left(\Delta-\frac{1}{\lambda}\right)}}_{=\beta_1}r.
\label{N12_gen}
\end{align}
Rational expectations requires the slope in \eqref{N12_gen} to coincide with the conjectured sensitivity $\beta_1$, which, after, reinjecting $\alpha_1$ and $\alpha_2$ from \eqref{N11_gen} into the above, yields $\Gamma^c(\beta_1^*)=0$ as defined in equation (21) in the main text. Further, this function satisfies:
\begin{eqnarray*}
\Gamma^{c}\!\Big(\frac{1}{\Delta}\Big)
&=& -\Big(e+\frac{k}{2}\Delta\Big)\frac{1}{\Delta^2}
+\Big(e\lambda-c\Delta\Big)\frac{1}{\Delta}
+c+\frac{k}{2}\lambda
=
\frac{(\Delta k+2e)(\Delta\lambda-1)}{2\Delta^2}>0,\\
\Gamma^{c}(\lambda)
&=&
-\Big(e+\frac{k}{2}\Delta\Big)\lambda^2
+\Big(e\lambda-c\Delta\Big)\lambda
+c+\frac{k}{2}\lambda
=
\frac{(1-\Delta\lambda)(2c+k\lambda)}{2}<0.
\end{eqnarray*}
Hence, there exists a unique positive root satisfying $\beta_1\in(1/\Delta,\lambda)$ and, given that the leading term of $\Gamma^c(.)$ is negative, it must hold that $(\Gamma^{c})'(\beta_1^*)<0$.$\Box$
\\ \medskip
\textbf{Proof of Proposition \ref{prop_FVy_equil}:}
We first derive the sender's optimal reporting strategy for a given conjectured belief function $P(r)=\beta_0+\beta_1 r$.
Taking the derivative of \eqref{FV} with respect to $r$ yields the first-order condition
\begin{equation}\label{N34_re}
\beta_1 y
- c\big(R(s,x,y)-s-x\big)
- e\beta_1\Big(P(R(s,x,y))-\lambda s\Big)=0,
\end{equation}
which implies the linear reporting rule
\begin{equation}\label{N31_re}
R(s,x,y)
=
\underbrace{-\frac{\beta_0 \beta_1 e}{c+\beta_1^2 e}}_{\equiv\alpha_0}
+\underbrace{\frac{c+\lambda \beta_1 e}{c+\beta_1^2 e}}_{\equiv\alpha_1}s
+\underbrace{\frac{c}{c+\beta_1^2 e}}_{\equiv\alpha_2}x
+\underbrace{\frac{\beta_1}{c+\beta_1^2 e}}_{\equiv\alpha_3}y.
\end{equation}

The receiver's posterior sensitivity is
\begin{equation}\label{beta1consFVx_re}
\beta_1
=
\frac{\mathrm{Cov}(\theta,r)}{\mathrm{Var}(r)}
=
\frac{\alpha_1}
{\frac{\alpha_1^2}{\lambda}
+\alpha_2^2\left(\Delta-\frac{1}{\lambda}\right)
+\alpha_3^2\frac{\sigma_y^2}{\sigma_\theta^2}}.
\end{equation}
Substituting the coefficients from \eqref{N31_re} into \eqref{beta1consFVx_re} yields the cubic equation \eqref{GammaFVy}. Further,
\[
\Gamma^{f}(0)
=
-\frac{c^2\sigma_\theta^4}{\lambda}<0<\Gamma^{f}(\lambda)
=
\sigma_\theta^2\left(
\sigma_y^2\lambda^2
+
c^2\sigma_\theta^2\Big(\Delta-\frac{1}{\lambda}\Big)
\right).
\]
Thus, there exists at least one root in $(0,\lambda)$.
Moreover,
\[
\frac{d^2\Gamma^{f}(\beta_1)}{d\beta_1^2}
=
\frac{6\sigma_\theta^2}{\lambda}\sigma_y^2\,\beta_1
+
\frac{2c\sigma_\theta^4}{\lambda}e
>0
\quad\text{for all }\beta_1\ge0,
\]
so $\Gamma^{f}$ is strictly convex on $[0,\infty)$, implying that the unique root $\beta_1^*$ is unique with $(\Gamma^{f})'(\beta_1^*)>0$. The intercept follows immediately from $\beta_0=-\beta_1\mathbb{E}[r]$.

The comparative statics follow from the implicit function theorem jointly with $(\Gamma^{f})'(\beta_1^*)>0$:

\begin{eqnarray*}
\frac{\partial \beta_1^*}{\partial e}&\propto& -\frac{\partial \Gamma^{f}(\beta_1^*)}{\partial e}=
\frac{c\sigma_\theta^4}{\lambda}\beta_1^*(\lambda-\beta_1^*)
>0\\
\frac{\partial \beta_1^*}{\partial \sigma_y^2}&\propto& -\frac{\partial \Gamma^{f}(\beta_1^*)}{\partial \sigma_y^2}=
-\frac{\sigma_\theta^2}{\lambda}(\beta_1^*)^3<0\\
\frac{\partial \beta_1^{*}}{\partial c}&\propto & -\frac{\partial \Gamma^{f}(\beta_1^*)}{\partial c}= 2\sigma_y^2(\beta_1^{*})^2
-
ce\,\sigma_\theta^2(\lambda-\beta_1^{*}).
\end{eqnarray*}
Therefore, $\beta_1^{*}$ is increasing in $c$ when $c$ is low and decreasing in $c$ when $c$ is high.$\Box$
\medskip\\
\textbf{Proof of Corollaries \ref{P31_FVy} and \ref{P31_FVy2}:}
From Proposition \ref{prop_FVy_equil},
\[
B
=
-\frac{\beta_0^*}{\beta_1^*}
=
m_x+\frac{\beta_1^*}{c}m_y.
\]
The comparative statics with respect to $m_x$ and $m_y$ are immediate. Further,
\[
\frac{\partial B}{\partial e}
=
\frac{m_y}{c}\frac{\partial \beta_1^*}{\partial e},
\qquad
\frac{\partial B}{\partial \sigma_y^2}
=
\frac{m_y}{c}\frac{\partial \beta_1^*}{\partial \sigma_y^2}.
\]
Since $\partial \beta_1^*/\partial e>0$ and $\partial \beta_1^*/\partial \sigma_y^2<0$, it follows that
$\partial B/\partial e>0$ and $\partial B/\partial \sigma_y^2<0$ when $m_y>0$.

Differentiating with respect to $c$ returns\[
\frac{\partial B}{\partial c}
=
\frac{m_y}{c^2}\left(c\frac{\partial \beta_1^*}{\partial c}-\beta_1^*\right)
=
-\frac{m_y}{c^2(\Gamma^{f})'(\beta_1^*)}
\left(
c\,\frac{\partial \Gamma^{f}(\beta_1^*)}{\partial c}
+\beta_1^*(\Gamma^{f})'(\beta_1^*)
\right).
\]
A direct calculation yields
\[
c\,\frac{\partial \Gamma^{f}(\beta_1^*)}{\partial c}
+\beta_1^*(\Gamma^{f})'(\beta_1^*)
=
\frac{c\sigma_\theta^4}{\lambda}\,(c+e\lambda\beta_1^*).
\]
Since $(\Gamma^{f})'(\beta_1^*)>0$, this implies
\[
\frac{\partial B}{\partial c}
=
-\frac{m_y}{c^2(\Gamma^{f})'(\beta_1^*)}
\frac{c\sigma_\theta^4}{\lambda}\,(c+e\lambda\beta_1^*)
<0
\qquad\text{for }m_y>0.
\]

We next turn to residual uncertainty. Using \eqref{varFVy},
\[
\Sigma
=
\sigma_\theta^2\left(1-\frac{\beta_1^*(c+e\lambda\beta_1^*)}{c+e(\beta_1^*)^2}\right).
\]

Differentiating with respect to $\beta_1^*$,

\[
\frac{\partial \Sigma}{\partial e}
=
-\sigma_\theta^2
\frac{c(\beta_1^*)^2(\lambda-\beta_1^*)}{(c+e(\beta_1^*)^2)^2}
-\sigma_\theta^2
\frac{c\big(c+e\beta_1^*(2\lambda-\beta_1^*)\big)}{(c+e(\beta_1^*)^2)^2}\frac{\partial \beta_1^*}{\partial e}
<0.
\]
The remaining comparative statics and the case of large $c$ follow readily. It remains to characterize the effect of $c$. For low $c$, retaining the leading term implies $\beta_1^*=b\sqrt{c}+o(\sqrt{c})$. Substituting this into \eqref{varFVy} then implies $\Sigma= Var(\theta)- \rho\sqrt{c}+ o(\sqrt{c})$
for some constant $\rho>0$. Hence $\partial \Sigma/\partial c<0$ when $c$ is low.$\Box$
\medskip\\
\textbf{Proof of Corollary \ref{wel2_FVy}:}
The sender's expected utility is
\[
W
=\mathbb{E}\big[yP(r)\big]
-\frac{c}{2}\mathbb{E}\big[(r-s-x)^2\big]
-\frac{e}{2}\mathbb{E}\big[(P(r)-\lambda s)^2\big].
\]

Using the linear reporting rule from Proposition \ref{prop_FVy_equil},
\[
\mathbb{E}[yP(r)]
=
\mathrm{Cov}(y,P(r))
=
\beta_1^*\alpha_3^*\sigma_y^2
=
\frac{(\beta_1^*)^2}{c+e(\beta_1^*)^2}\sigma_y^2.
\]

Next, $\mathbb{E}[r-s-x]
=
\frac{\beta_1^*}{c}m_y$ and a direct substitution using the coefficients in Proposition \ref{prop_FVy_equil} returns
\begin{eqnarray*}
\mathbb{E}\big[(r-s-x)^2\big]
&=&
\frac{(\beta_1^*)^2m_y^2}{c^2}
+
\frac{(\beta_1^*)^2\left[e^2\sigma_\theta^2\big(\Delta(\beta_1^*)^2-2\beta_1^*+\lambda\big)+\sigma_y^2\right]}
{(c+e(\beta_1^*)^2)^2}\\
\mathbb{E}\big[(P(r)-\lambda s)^2\big]
&=&
\frac{c^2\sigma_\theta^2\big(\Delta(\beta_1^*)^2-2\beta_1^*+\lambda\big)+(\beta_1^*)^4\sigma_y^2}
{(c+e(\beta_1^*)^2)^2}.
\end{eqnarray*}
Substituting these expressions into $W$ and simplifying yields
\begin{eqnarray*}
W
&=&
-\frac{
\Delta(\beta_1^*)^2c^2e\sigma_\theta^2
+(\beta_1^*)^4 e m_y^2
+(\beta_1^*)^2 c m_y^2
-(\beta_1^*)^2 c \sigma_y^2
-2\beta_1^* c^2 e \sigma_\theta^2
+c^2 e \lambda \sigma_\theta^2}
{2c\big(c+e(\beta_1^*)^2\big)}\\
&=&
\underbrace{-\frac{e\lambda\sigma_\theta^2}{2}}_{W^{pool}}
+
\frac{\beta_1^*}{2}\left(
e\sigma_\theta^2+\frac{\beta_1^*}{c}\big(\sigma_y^2-m_y^2\big)
\right).
\end{eqnarray*}

It follows immediately that $W$ is decreasing in $|m_y|$.
If $m_y=0$, then
\[
W
=
-\frac{e\lambda\sigma_\theta^2}{2}
+
\frac{\beta_1^*}{2}\left(
e\sigma_\theta^2+\frac{\beta_1^*}{c}\sigma_y^2
\right)=-\frac{c\sigma_\theta^2}{2\beta_1^*}\,(\Delta\beta_1^*-1).
\]
By convexity of $\Gamma^f(.)$ and the fact that the root is unique,
$\Delta \beta_1^*-1<0$ is equivalent to \[
\Gamma^f(1/\Delta)
=
\frac{\sigma_\theta^2}{\Delta^3\lambda}
\Big(
\sigma_y^2
-
ce\,\sigma_\theta^2\Delta(\Delta\lambda-1)
\Big)>0,
\]
which is equivalent to $\sigma_y^2>ce\,\sigma_\theta^2\Delta(\Delta\lambda-1)$. Finally, if $|m_y|\le \sigma_y$, 
\[
W-W^{pool}
=
\frac{\beta_1^*}{2}\left(
e\sigma_\theta^2+\frac{\beta_1^*}{c}\big(\sigma_y^2-m_y^2\big)
\right)
>0,
\]
so the sender strictly prefers the informative LREE to pooling.
\hfill$\Box$
\\ \medskip
\textbf{Proof of Corollary \ref{JU_FVy}:}
We first show that for the model to converge to full-revealing messages, $e_i$ must diverge. In any such limit, $\alpha_{2,i}$ and $\alpha_{3,i}$ must converge to zero which, using their equilibrium coefficients, implies

\[
\alpha_{2,i}
=
\frac{c_i}{c_i+e_i(\beta_{1,i}^*)^2},
\qquad
\alpha_{3,i}
=
\frac{\beta_{1,i}^*}{c_i+e_i(\beta_{1,i}^*)^2}.
\]
Hence
$e_i(\beta_{1,i}^*)^2\to\infty$,
which requires  $e_i\to\infty$.

Similar steps as the baseline model then show that, for any sequence $\omega$ with $c_i \rightarrow c^0$, the sender's asymptotic expected utility is 
\[
W_\infty(\omega)
=
-\frac12\left(
c^0\sigma_x^2+\frac{\lambda^2 m_y^2}{c^0}
\right),
\]
which is maximized at $c_\infty^{**}=\frac{\lambda m_y}{\sigma_x}$ and with maximum value $-\lambda m_y\sigma_x$.
$\Box$
\\ \medskip
\textbf{Proof of Proposition \ref{LMB}:} In the modified game \eqref{U11b}, the sender's payoff is
$u(r;s,x)=P(r)-\frac{c}{2}(r-s-x)^2-\frac{e}{2}\big(P(r)-\mathbb{E}(\theta| s+x,r)\big)^2$.
For any linear conjecture $P(r)=\beta_0+\beta_1 r$, the posterior
$\mathbb{E}(\theta| s+x,r)$ is a function of $(s+x,r)$ only (since $\theta$, $s+x$, and $r$
are jointly Gaussian and the conditioning variables are $s+x$ and $r$).
The lying cost $(r-s-x)^2$ likewise depends on $(s,x)$ only through $s+x$.
Therefore the sender's objective at any $(s,x)$ is a function of $r$ and the
sufficient aggregate $s+x$ alone. Because the objective is strictly concave
in $r$ (the second derivative is $-(c+e\beta_1^2)<0$), the pointwise best response
$r^*(s,x)$ is unique and depends on $(s,x)$ only through $s+x$.
Consequently, any linear best response $R(s,x)=\alpha_0+\alpha_1 s+\alpha_2 x$
must satisfy $\alpha_1=\alpha_2$.$\Box$

\noindent\textbf{Proof of Proposition \ref{prop:iter_convergence}:} Fix $n$ and suppose that $P_n(r)=\beta_{0,n}+\beta_{1,n}r$ with $\beta_{1,n}\ge 0$. The first-order condition yields the unique best response $R_n(s,x)=\alpha_{0,n}+\alpha_{1,n}s+\alpha_{2,n}x$,
with coefficients given by \eqref{eq:alpha_iter}. Since $(\theta,s,x)$ are jointly Gaussian and $R_n(.)$ is linear, the posterior expectation $P_{n+1}(r)=\mathbb{E}[\theta| R_n(s,x)=r]$ is also linear, with coefficients given by \eqref{eq:beta_update_general}.

\smallskip
\noindent 

Substituting \eqref{eq:alpha_iter} into  \eqref{eq:beta_update_general} returns $\beta_{1,n+1}=\mathcal{T}(\beta_{1,n})$, where $\mathcal{T}$ is defined by \eqref{eq:Tdef_global}. Note that
\[
\lambda\Delta-1=\frac{\sigma_x^2}{\sigma_\theta^2+\sigma_\epsilon^2}>0,
\]
so the denominator of $\mathcal{T}$ is strictly positive for every $\beta\ge 0$.
Re-arranging,\begin{equation}\label{eq:T_minus_id_global}
\mathcal{T}(\beta)-\beta
=
\frac{c\lambda}{(c+e\lambda\beta)^2+c^2(\lambda\Delta-1)}\;\Gamma(\beta).
\end{equation}
Because the denominator is positive, $\mathcal{T}(\beta)-\beta$ has the same sign as $\Gamma(\beta)$.

\smallskip
\noindent 

From Proposition 1 in the main text, $\Gamma$ has a unique positive root $\beta_1^*\in(0,\lambda)$. Rewriting $\Gamma(\beta_1^*)=0$ as $c\Delta=e\lambda+\frac{c}{\beta_1^*}-e\beta_1^*$ and substituting it into \eqref{eq:Tdef_global} yields
\begin{equation}\label{eq:T_minus_star}
\mathcal{T}(\beta)-\beta_1^*
=
\frac{e\lambda(\beta-\beta_1^*)\big[c(\lambda+\beta-\beta_1^*)+e\lambda\beta^2\big]}
{(c+e\lambda\beta)^2+c^2(\lambda\Delta-1)}.
\end{equation}
The denominator is strictly positive. The bracketed term is also strictly positive for every $\beta\ge 0$ because $\lambda-\beta_1^*>0$. Therefore
$\mathcal{T}(\beta)-\beta_1^*\propto \beta-\beta_1^*$, which implies that the $\mathcal{T}(\beta)$ cannot cross $\beta_1^*$. It is then readily verified that $\mathcal{T}^n(\beta)$ monotonically increases (decreases) to $\beta_1^*$ if $\beta<\beta_1^*$\ ($\beta>\beta_1^*$). Convergence of the  rest of the equilibrium coefficients follow immediately. $\Box$

\medskip
\textbf{Proof of Proposition \ref{prop:B5_onpath}:}
Given a report $r_0\in\mathbb{R}$. Define
\[
s_0\equiv \frac{P(r_0)-1/e}{\lambda},
\qquad
\textrm{with } x_0\equiv r_0-s_0.
\]

For this type,  $\lambda s_0=P(r_0)-\frac{1}{e}$ and $s_0+x_0=r_0$. Therefore, the sender's utility from an arbitrary report $r$ is
\[
u(r;s_0,x_0)
=
P(r)-\frac{c}{2}(r-r_0)^2-\frac{e}{2}\Big(P(r)-P(r_0)+\frac{1}{e}\Big)^2.
\]
The term
\[
p-\frac{e}{2}\Big(p-P(r_0)+\frac{1}{e}\Big)^2
\]
is strictly concave quadratic in $p$ with derivative
\[
1-e\Big(p-P(r_0)+\frac{1}{e}\Big)=e\big(P(r_0)-p\big),
\]
so it is uniquely maximized at $p=P(r_0)$. Therefore, among all reports, the report $r_0$ maximizes the posterior term. In addition, if $r\neq r_0$, the lying term $-\frac{c}{2}(r-r_0)^2$ is strictly negative because $c>0$. Therefore $u(r;s_0,x_0)<u(r_0;s_0,x_0)$ for every $r\neq r_0$ and it follows that $r_0$ is the unique optimal report for the type $(s_0,x_0)$. Since $r_0$ was arbitrary, every report in $\mathbb{R}$ is used on path.
An admissible equilibrium cannot therefore contain an interval with only finitely many on-path reports, so that partitional intervals are impossible and every admissible equilibrium is continuous. $\Box$

\medskip

\noindent\textbf{Proof of Proposition \ref{prop:B5_unique}:}
By Proposition~\ref{prop:B5_onpath}, any admissible equilibrium is continuous. Consider such an equilibrium and suppose that it is regular.

We first show that the sender's best-response map is continuous. The utility function in equation (1) in the main text is continuous in $(r,s,x)$, and regularity requires the maximizer $R(s,x)$ to be unique for every type. Berge's maximum theorem therefore implies that $R(.)$ is continuous. On any interval where $P(.)$ is $C^2$, the strict second-order condition in Definition~\ref{def:B5_regular}(iii) allows us to apply the implicit function theorem, so $R(.)$ is locally $C^1$ on such an interval.

Now consider a report $r$ lying in the interior of one of the smooth intervals of Definition~\ref{def:B5_regular}(i). For any type $(s,x)$ with $R(s,x)=r$, the first-order condition from equation (1) in the main text can be rearranged as
\begin{equation}
\label{eq:B5_line}
\big(e\lambda P'(r)+c\big)s+c x = c r-P'(r)+eP(r)P'(r).
\end{equation}
 \eqref{eq:B5_line} is the equation of a straight line in the $(s,x)$-plane. Thus, locally, the types choosing report $r$ must lie on that line. Because $R(.)$ is continuous, the set of types choosing $r$ is closed in that line. Because $R(.)$ is locally $C^1$ and the second-order condition is strict, the implicit function theorem implies that this set is also open in that line. Since the line is connected and nonempty, the set of types choosing $r$ must coincide with the entire line.

Therefore each on-path report in a smooth interval is associated with a full straight line of types. Distinct reports cannot be chosen by the same type because the optimum is unique. Hence the full lines associated with two distinct reports cannot intersect. But two distinct full lines in the plane can fail to intersect only if they are parallel. It follows that the slope in \eqref{eq:B5_line} must be the same for every report in every smooth interval. Since the coefficient on $x$ is the constant $c$, the coefficient on $s$ must also be constant. Therefore $P'(r)$ is constant on every smooth interval.
Because there are only finitely many smooth intervals and $P(.)$ is continuous across their boundaries, the same constant slope must hold everywhere, implying that $P(.)$ is affine on $\mathbb{R}$.$\Box$

\end{singlespace}

\clearpage
\bibliographystyle{chicago}
\renewcommand{\refname}{References}
\bibliography{paper_specific_references}
\end{document}